# Performance evaluation of a production line operated under an echelon buffer policy

George Liberopoulos

Department of Mechanical Engineering, University of Thessaly, Volos 38334, Greece, e: glib@mie.uth.gr

**Abstract**

We consider a production line consisting of several machines in series separated by intermediate finite-capacity buffers. The line operates under an "echelon buffer" (EB) policy according to which each machine can store the parts that it produces in any of its downstream buffers if the next machine is occupied. If the capacities of all but the last buffer are zero, the EB policy is equivalent to CONWIP. To evaluate the performance of the line under the EB policy, we model it as a queueing network, and we develop a method that decomposes this network into as many nested segments as there are buffers and approximates each segment with a two-machine subsystem that can be analyzed in isolation. For the case where the machines have geometrically distributed processing times, we model each subsystem as a two-dimensional Markov chain that can be solved numerically. The parameters of the subsystems are determined by relationships among the flows of parts through the echelon buffers in the original system. An iterative algorithm is developed to solve these relationships. We use this method to evaluate the performance of several instances of 5- and 10-machine lines including cases where the EB policy is equivalent to CONWIP. Our numerical results show that this method is highly accurate and computationally efficient. We also compare the performance of the EB policy against the performance of the traditional "installation buffer" policy according to which each machine can store the parts that it produces only in its immediate downstream buffer if the next machine is occupied. Supplementary materials are available for this article. Go to the publisher's online edition of *IISE Transactions*, datasets, additional tables, detailed proofs, etc.

# 1   Introduction

Production lines are the prevailing layout in high-volume discrete-part manufacturing. A production line consists of several machines that are visited by all parts once and in a fixed sequence. The time that a part spends on a machine is often random because of unpredictable disruptions in the production process and/or variability in the processing requirements of the part. This randomness causes congestion in the line and adversely affects system performance, most notably throughput. One way to increase throughput is to raise the processing rates of the machines starting with the slowest one. Another way is to reduce the variance of processing times causing the congestion. Both approaches require the adoption of good engineering and operating practices at the machine level and possibly investing in new equipment. A third alternative is to reduce the effect of randomness by inserting finite buffers between the machines so that parts flow from machine to buffer to machine and so on until they exit the line. Inserting a buffer between two machines speeds up the line by decoupling the operation of the machines, as long as this buffer is neither full nor empty, hence limiting the propagation of processing time delays.

In the traditional way of operating a line with intermediate buffers, each machine is allowed to store the parts that it produces only in its immediate downstream buffer if the next machine is occupied. We refer to the ensemble of that buffer and the next machine as installation buffer, and to the resulting policy as installation buffer (IB) policy. Under the IB policy, a machine is blocked from processing a part if the number of parts that have been produced by it but have not yet exited the next machine, which is referred to as the installation work-in-process (WIP) following the machine, is equal to the capacity of the installation buffer downstream of the machine.

Inserting buffers between machines comes at a cost of additional WIP inventory, capital investment, and floor space. Depending on the industry, such a cost can be quite high. For instance, in a car manufacturing body shop, the production and material cost of a part may go up to $10K, based on discussions with experts. With a 5% interest rate, the annual unit cost of WIP can be as high as $500. The capital investment for a single buffer space in a body shop can cost thousands of dollars (see Askin and Fowler 2013 summarizing Lagershausen et *al*. 2013); an investment of $5-10K per space, resulting in an annual capital cost of $250-500 at 5% interest rate, is not unusual. On top of this cost, one may have to account for depreciation over the lifetime of the buffer space that could be up to 10 years. With a linear depreciation, this could add another $500-1,000 to the annual cost of the investment.

The optimal allocation of storage capacity among the intermediate buffers is one of the most widely studied problems in manufacturing systems research. Even if the total capacity has been optimized, storing parts locally in the buffers immediately following the machines does not take full advantage of this capacity. When the cost of buffer space is significant, it may be worthwhile to consider increasing the utilization of the existing buffers before setting out to increase total buffer capacity. One way to increase buffer utilization is to allow the machines to store the parts that they produce in buffers other than their immediate downstream buffers. Such a mode of operation is not unusual in practice, especially in systems where material handling is performed by humans. We have witnessed this in a producer of large industrial conveyor belts where operators transfer WIP material with forklift trucks, in a manufacturer of medium-sized metallic parts where workers transfer parts with trolleys, and in other production environments. The question is, can this be done systematically, and if so, what are the gains and losses in performance?

In this paper, we consider a policy aimed at increasing the utilization of buffers in a production line by allowing each machine to store the parts that it produces in any of its downstream buffers if the next machine is occupied, rather than only in its immediate downstream buffer, as is the case under the classical IB policy. The ensemble of all the downstream buffers and the next machine is referred to as echelon buffer, and the resulting policy is referred to as echelon buffer (EB) policy. Under the EB policy, a machine is blocked from processing a part if the number of parts that have been produced by it but have not yet exited the line, which is referred to as the echelon WIP following the machine, is equal to the capacity of the echelon buffer downstream of the machine.

The terms "installation" and "echelon" originate from inventory control theory, with the term echelon dating back to Clark (1958). In a multi-stage inventory system, under an "installation stock" policy, the ordering decision at each stage (installation) is based on the inventory position at this stage, whereas under an "echelon stock" policy, it is based on the echelon stock position, which is defined as the sum of the installation stock positions at this stage and all its downstream stages. Axsäter and Rosling (1993) show that for reorder point-reorder quantity rules, echelon stock policies are in general superior to installation stock policies. Given that results for inventory systems do not generally carry over to production systems, because the limited capacity in the latter systems causes congestion and affects production time in a non-trivial way, the ultimate goal of this research is to compare the performance of the EB policy against that of the IB policy in the case of production lines.

To evaluate the performance of a production line under the EB policy, we model it as a queueing network, and we develop a method that decomposes this network into as many nested segments as there are buffers and approximates each segment with a two-machine subsystem that can be analyzed in isolation. For the case where the machines have geometrically distributed processing times, we model each subsystem as a two-dimensional Markov chain that can be solved numerically. The parameters of the subsystems are determined by relationships among the flows of parts through the echelon buffers in the original system. An iterative algorithm is developed to solve these relationships. We evaluate the accuracy of this method by comparing it against simulation for several instances of 5- and a 10-machine lines, including cases where the EB policy is equivalent to CONWIP. Our numerical results show that the method is very accurate and computationally efficient. We also compare the performance of the EB policy against the performance of the IB policy which we evaluate by simulation. The use this method to optimally design the echelon buffer capacities is deferred for future consideration.

The remainder of this paper is organized as follows. In Section 2, we describe the operation of a production line under the EB policy and discuss some of its advantages and disadvantages. In Section 3, we review the related literature. In Section 4, we describe the discrete-time queuing network model of the EB-controlled line. In Section 5, we present the decomposition-based approximation method for analyzing this network. In Section 6, we present the analysis of each subsystem of the decomposition, and in Section 7, we present the analysis of the entire system. In Section 8, we present numerical results on the performance of the decomposition method and on the effect of system parameters on performance. We also compare the performance of EB against the performance of IB. Finally, we draw conclusions in Section 9. Certain derivations and numerical results are included in an online supplement to this paper.

## 2 Description of a production line operated under an EB policy

We consider a production line consisting of $N$ machines in series, denoted by $M_n, n = 1, \ldots, N$, and $N - 1$ intermediate finite-capacity buffers, denoted by $B_n, n = 1, \ldots, N - 1$. Each machine has unit capacity, while the capacity of buffer $B_n$ is denoted by $C_n$, where $C_n \geq 0, i = 1, \ldots, N - 1$. Under the EB policy introduced in the previous section, any part produced by machine $M_n$ is stored in the echelon buffer following $M_n$. That buffer is denoted by $B_n^E$ and is defined as $B_n^E = B_n \cup \ldots \cup B_{N-1} \cup M_{n+1}, i = 1, \ldots, N - 1$. In addition, $M_n$ is blocked from processing a part if the echelon WIP following it (i.e., the number of parts that have been produced by it but have not yet exited the line) is equal to the capacity of $B_n^E$, which is given by $1 + \sum_{m=n}^{N-1} C_m$. This implies that the cap on the total line WIP following $M_1$ is $1 + \sum_{n=1}^{N-1} C_n$.

In contrast, under the traditional IB policy, any part produced by machine $M_n$ can be stored only in the installation buffer following $M_n$. That buffer is denoted by $B_n^I$ and is defined as $B_n^I = B_n \cup M_{n+1}, i = 1, \ldots, N - 1$. Moreover, $M_n$ is blocked from processing a part if the installation WIP following it (i.e., the number of parts that have been produced by it but have not yet departed from the next machine, $M_{n+1}$) is equal to the capacity of $B_n^I$, which is given by $1 + C_n$. In this case, the maximum total line WIP, which is defined as the WIP following $M_1$, is $\sum_{n=1}^{N-1}(1 + C_n) = N - 1 + \sum_{n=1}^{N-1} C_n$.

The type of blocking that we have considered here is known to as blocking before service with position occupied (BBS-PO) and is presumed in Gershwin and Berman's (1981) seminal two-machine model and in several other works. Another blocking mechanism that is often encountered in manufacturing is blocking after service (BAS). The analysis under both mechanisms is similar, and the difference in performance between them becomes inconsiderable for large buffer sizes (Dallery and Gershwin, 1992). Throughout this paper, we adopt the BBS-PO convention because it leads to a simpler description.

Figure 1 shows a production line operated under an EB policy for $N = 4$. To further clarify how the EB policy works, we note that machine $M_1$ stores the parts that it produces in buffers $B_1, B_2$, or $B_3$ if $M_2$ is occupied, with $B_1$ having the highest priority and $B_3$ the lowest; hence, $B_1^E = B_1 \cup B_2 \cup B_3 \cup M_2$. Similarly, machine $M_2$ stores the parts that it produces in buffers $B_2$ or $B_3$ if $M_3$ is occupied. Finally, machine $M_3$ stores the parts that it produces in buffer $B_3$ if $M_4$ is occupied. Furthermore, $M_1$ is blocked from processing a part if the number of parts that have been produced by it but have not yet exited the line is equal to the capacity of $B_1^E$ which is $1 + C_1 + C_2 + C_3$. Similarly, $M_2$ is blocked if the number of parts that have been produced by it but have not yet exited the line is equal to $1 + C_2 + C_3$. Finally, $M_3$ is blocked if the number of parts that have been produced by it but have not yet exited the line is equal to $1 + C_3$.

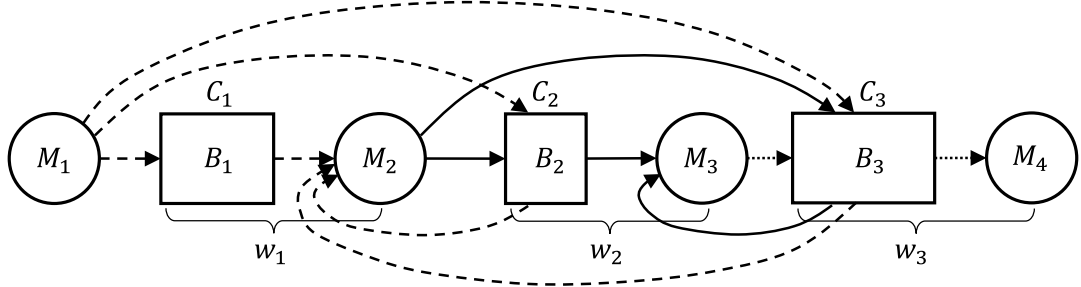

**Figure 1.** Production line with intermediate finite buffers operated under an EB policy.

If the physical layout of the production line is one where there is a common storage area on the side of the line rather than separate buffers between the machines, then under the EB policy, this area is divided into compartments that play the same role as the intermediate buffers in the traditional serial layout. In this case, the flow of parts is identical to that in the serial layout shown in Figure 1, except that the buffers are clustered compartments, as is shown in Figure 2(a).

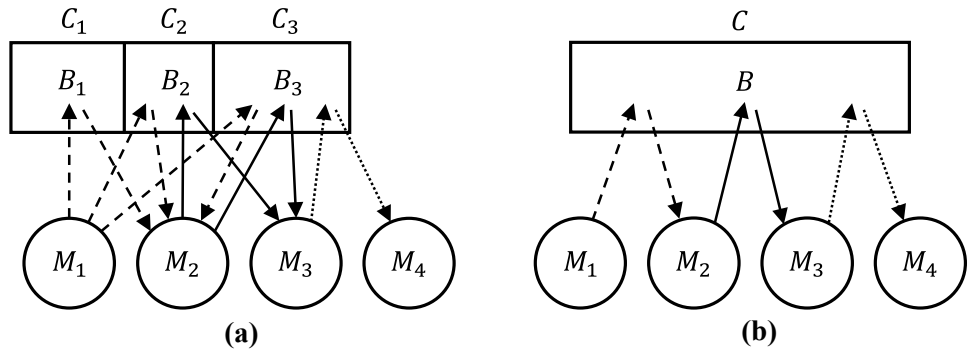

**Figure 2.** Production line with a common storage area divided into (a) several compartments or (b) a single compartment, operated under an EB policy.

If the capacities of all intermediate buffers, except possibly the last one, are zero, i.e., $C_1 = C_2 = 0$ and $C_3 = C \geq 0$, then every machine stores the parts that it produces in the last and only buffer if the next machine is occupied. This is shown in Figure 2(b), where the last buffer is denoted by $B$ and is drawn on the side of the line as a common buffer. In this case, the first machine is blocked from processing a part if the number of parts that have been produced by it but have not yet exited the line is equal to $1 + C$. No other machine is ever blocked. This way of operation is identical to the operation of a CONWIP system, where parts are not allowed to be released into the system if the total WIP is at the WIP-cap. Note that here we have adopted the original definition of CONWIP according to which the WIP in the system is capped rather than constant (Spearman et *al*. 1990), even though CONWIP stands for "*con*stant work *in p*rocess". For the purposes of this paper, we will henceforth refer to an EB policy where all buffers except the last one have zero capacities and the last buffer has capacity $C \geq 0$, as CONWIP with WIP-cap $1 + C$.

Note that a line of three of more machines operated under EB has a lower total WIP-cap than the same line operated under IB. More specifically, for an $N$-machine line ($N \geq 3$), the total WIP-cap under EB is $1 + \sum_{n=1}^{N-1} C_n$, whereas under IB it is $N - 1 + \sum_{n=1}^{N-1} C_n$, as was mentioned earlier. The difference in WIP-cap between the two policies is significant only for short lines with very low buffer capacities. As a result of this difference, in such lines, it may happen that the EB policy yields lower average throughput and/or WIP than the IB policy does. In general, however, it is expected that a production line operated under an EB policy should have higher average throughput – at the cost of higher average WIP – than the same line operated under an IB policy, because of the higher utilization of buffer space under EB.

An important advantage of the EB policy, besides increasing buffer space utilization, is that it uses global information because it enables each machine to process parts based on the entire echelon WIP level downstream of this machine. This can be beneficial especially if the WIP holding cost increases significantly downstream the line, as is the case with products that have high added value. In contrast, the IB policy uses only local information because it enables each machine to process parts based on the WIP level of the local installation buffer immediately following it.

A shortcoming of the EB policy is that it has increased material handling requirements compared to the IB policy. Modern technology, however, can handle such increased requirements at affordable costs (Matta et *al*. 2005). Today, there exist several affordable modular and reconfigurable material handling solutions that are less automated than traditional systems and can be assembled in a flexible way to transport parts in the manufacturing floor (Furmans et *al*. 2010). Many of the material handling ideas and equipment that are used today in production lines with reentrant flows can also be used to implement the EB policy. The interested reader is referred to two surveys on automated material handling systems in semiconductor manufacturing by Agrawal and Heragu (2006) and Montoya-Torres (2006). The material handling technology for implementing the EB policy can also be found in classical flexible manufacturing systems and their successors, reconfigurable manufacturing systems, where typically pallets are sent back and forth to the work centers. Such movements of material are sometimes referred to as backtracking and bypassing and have been extensively studied in the context of the more general facility layout problem (e.g., see the reviews by Hassan 1994 and Drira et *al*. 2007). Finally, the related problem of controlling the flow of automated guided vehicles in manufacturing environments with complex flows has also been studied extensively. Two surveys on this issue are Le-Anh and De Koster (2006) and Vis (2006).

Another issue is distinguishing parts that are stored in the same buffer but are in different stages of their processing. Here, optical or electronic means can be used. Schuler and Darabi (2016) describe a case of a manufacturing facility producing mobile devices where parts in neighboring stages of their production are manually stored in shared buffer clusters. When a part is picked up for processing by a machine, its bar code or RFID is scanned to ensure that the preceding operations have been completed. When the part is processed by the machine, the operator scans the part to inform the information system that it has been processed by this machine, before putting it back to storage. While the machine is processing the part, the system identifies the next part to be picked up for processing via RFID and notifies the operator via a light or some other indicator.

## 3    Literature review

The role of intermediate storage buffers in mitigating the adverse effect of process time variability on the efficiency of manufacturing flow lines has been researched for over fifty years. Buzacott (1967, 1971) are among the earliest references in English in this topic. In the years that followed, the analysis of flow lines rapidly evolved into a thriving research field with significant practical implications. Many of the core ideas and methods were developed by the 1990's and appeared in numerous papers, surveys, and books (e.g., Dallery and Gershwin 1992; Buzacott and Shanthikumar 1993; Askin and Standridge 1993; Tempelmeier and Kuhn 1993; Papadopoulos et *al*. 1993; Gershwin 1994; Papadopoulos and Heavey 1996; Altiok 1997). Since then, improvements, extensions, and generalizations of previously defined problems were established, and new problems and solution methodologies were developed. A recent overview and a textbook on the subject are Li and Meerkov (2009) and Li et *al*. (2009). Most of the issues that have been studied throughout these years fall into one of three categories: (1) modeling aspects, (2) performance evaluation, and (3) optimization. A recent literature review concerning these three dimensions can be found in Shi (2012). In the next two paragraphs, we briefly review categories (1) and (2). We defer the review of category (3) to future work where we plan to use the method developed in this paper to optimally design the EB policy and compare it against other policies.

Most of the modeling aspects of production lines have been covered in Dallery and Gershwin (1992). These aspects concern the stochastic nature of machine processing times, blocking mechanisms, the nature of material flow and time (continuous/discrete), etc. The simplest way of capturing the stochastic nature of machine processing times is to model them as geometrically distributed random variables. Such machines are often referred to as following the Bernoulli reliability model or simply as Bernoulli machines (Li and Meerkov, 2009). The Bernoulli machine model has been used routinely for studying various aspects of production lines (e.g., Billier et *al*. 2009; Li and Meerkov 2000; Meerkov and Zhang 2008, 2011). Its continuous-time equivalent, the exponential processing time model, has also been used extensively in the literature (e.g., Altiok 1997; Li and Meerkov 2009). In this paper, we adopt the Bernoulli machine model.

As far as the performance evaluation of production lines is concerned, many different techniques have been invoked, including simulation, Markov chain analysis, approximate analytical methods, and decomposition methods, among others. Decomposition methods in particular are two-step approaches that are based on decomposing long lines of many machines and intermediate buffers into several smaller tractable building blocks – usually two-machine, one-buffer pseudo-lines. The buffer in each pseudo-line represents one of the intermediate buffers in the original line. Typically, in the first step of such a method, the performance of each two-machine pseudo-line is evaluated given the parameters of the two machines. Many different models of two-machine systems have been analyzed in the literature, starting from the earlier simpler models (e.g., Buzacott 1967; Gershwin and Berman 1981) and advancing to more complex and general models in recent years (e.g., Tan and Gershwin 2009). In the second step, the parameters of the two-machine pseudo-line are determined by relationships among the flows of parts through the intermediate buffers of the original system. The literature on decomposition methods is extensive, spanning several decades (e.g., Gershwin 1987; Dallery et *al*. 1988; Levantesi et *al*. 2003; Colledani and Gershwin 2013). Most of the decomposition methods that have been developed concern production lines operated under the traditional IB policy. Under that policy, parts move unidirectionally from upstream to downstream buffers; hence, the decoupling effect of each buffer is clear. Under the EB policy, however, the decoupling effect is more complex because parts may also move in the opposite direction from downstream to upstream buffers. To address this complexity, special attention is required.

Finally, we note that the concept of temporarily storing parts in shared buffers when the intermediate dedicated buffers following the machines are full, though used in practice, has not been thoroughly investigated in the literature. Tempelmeier et *al*. (1989) (and later Tempelmeier et *al*. 1993) is one of the first attempts to model a flexible manufacturing system (FMS) with some sharing of buffer space. The FMS consists of several workstations. Each workstation has one or more machines and a local finite buffer. A central buffer is also available for storing parts if there is no space in the local buffers. Parts are mounted onto pallets that come in a fixed number. To evaluate the performance of the system, they model the FMS as a closed queueing network (CQN) with blocking and solve it using numerical approximation techniques (Bruell and Balbo 1978).

In a related study, Matta et *al*. (2005) consider a closed flow line with finite dedicated intermediate buffers and a finite shared common buffer that can be used by any machine whose dedicated buffer is full. It takes a certain travel time to move parts from the dedicated buffers to the common buffer and vice versa.

This time, if long, may cause the machines to starve. For 5-machine lines, they evaluate the throughput rate under different dedicated and shared buffer allocation configurations using simulation. They also discuss useful practical technological and economic considerations concerning the implementation of the shared buffers in real flow lines. In this paper, we assume that the time to transfer parts from the machines to remote buffers and back is negligible compared to the processing times on the machines. Even if this time is not negligible, however, it is still possible to neglect it by carefully planning the transfer of parts from remote buffers to the machines before these machines runs out of parts from their local buffers.

There have been several other applications of CQN modeling and analysis to manufacturing, and in particular kanban and other pull control mechanisms (e.g., Di Mascolo et *al*. 1996; Baynat et *al*. 2001; Satyam and Krishnamurthi 2008), as well as production lines with finite buffers (e.g., Lagershausen et *al*. 2013). In one of these applications, Koukoumialos and Liberopoulos (2005) develop an analytical approximation method for the performance evaluation of a multi-stage production inventory system operated under an echelon kanban (EK) policy. The connection between the EK policy and the EB policy becomes evident once the association between the number of available echelon kanbans in EK and the number of available buffer spaces in EB is made. In the EK system, each stage has an input buffer and is an open queueing network of machines with load-dependent continuous-time processing rates. The main production unit in this paper, on the other hand, is a Bernoulli machine with no input buffer. As a result of this difference, in the EK system, blockages of parts happen at output buffers rather than on machines. Moreover, in the EK system, the analysis of each subsystem in isolation involves a product-form approximation technique for solving a CQN problem, whereas in the EB system, each subsystem is evaluated using exact discrete-time Markov chain analysis. As a result, the accuracy of the decomposition method is higher in the EB system than it is in the EK system.

Finally, Zhou and Lian (2011) consider a 2-stage tandem network where each stage has a single exponential server. Customers arrive to the first stage following a Poisson process. The waiting customers in the two stages share all or part of a common finite buffer. They model the system as a two-dimensional Markov chain and compute the stationary probability distribution and the sojourn time distribution. They also present limited results on the shared buffer size that minimizes total buffer costs subject to minimum customer loss probability and maximum waiting time constraints. Their model, although limited to two servers, is somewhat similar to ours if one considers the external arrivals as being generated by a machine.

## 4 Model of a production line operated under an EB policy

In this section, we develop a queueing network model of a production line operated under an EB policy. This model is denoted by $L$ and consists of the $N$ machines of the line, $M_1, M_2, \ldots, M_N$, separated by $N-1$ infinite-capacity buffers, denoted by $Q_1, Q_2, \ldots, Q_{N-1}$. Figure 3 displays the queueing network model of the 4-machine line shown in Figure 1.

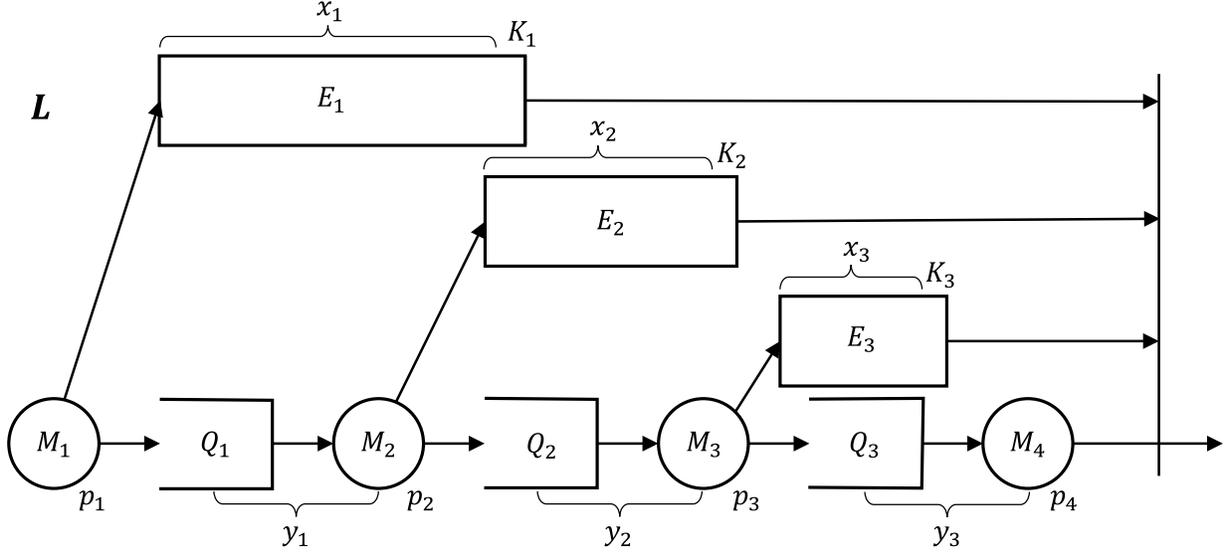

**Figure 3.** Queueing network model of the production line operated under EB shown in Figure 1.

For a general $N$-machine line model, we make the following assumptions:

(1) Parts flow from outside the system to $M_1$ to $Q_1$ to $M_2$ to … to $Q_{N-1}$ to $M_N$ and exit the system.
(2) Time is divided in discrete, equal-length periods.
(3) In each period, $M_n$ produces a part with probability $p_n$ unless it is starved or blocked, $n = 1, \ldots, N$. This implies that the processing time of a part on machine $M_n$ is geometrically distributed with mean $1/p_n$, variance $(1 - p_n)/p_n^2$, and squared coefficient of variation $1 - p_n$. Probability $p_n$ is referred to as the production probability (or rate) of machine $M_n$ in isolation.
(4) The number of parts in buffer $Q_n$ and machine $M_{n+1}$ is denoted by $y_n$ and is referred to as the stage WIP following machine $M_n, n = 1, \ldots, N - 1$; hence, $Q_n$ is referred to as the stage buffer following $M_n$. Note that $y_n$ is a function of the discrete time, but we omit this dependence for notational simplicity.
(5) When a part flows from machine $M_n$ to buffer $Q_n$, a token is generated and is placed in an associated finite buffer denoted by $E_n, n = 1, \ldots, N - 1$. This token is removed from $E_n$ and is discarded when the part exits the last machine, $M_N$. The total number of tokens in $E_n$ is denoted by $x_n$. Clearly, $x_n$ is equal to the number of parts that have been produced by machine $M_n$ but have not yet exited the network, i.e., it is equal to the echelon WIP following machine $M_n$ in the physical line. It is easy to see that $x_n$ is also equal to the sum of the stage WIP levels downstream of $M_n$, i.e.,

$$x_n = \sum_{m=n}^{N-1} y_m, \quad n = 1, \ldots, N - 1. \tag{1}$$

Note that $x_n$, like $y_n$, is a function of the discrete time but we omit this dependence for notational simplicity. Expression (1) can also be written recursively as

$$x_n = y_n + x_{n+1}, \quad n = 1, \ldots, N-2; \quad x_{N-1} = y_{N-1}. \tag{2}$$

(6) The capacity of buffer $E_n$ is denoted by $K_n$ and is equal to the capacity of echelon buffer $B_n^E$, $n = 1, \ldots, N-1$, in the physical line, i.e.,

$$K_n = 1 + \sum_{m=n}^{N-1} C_m > 0, n = 1, \ldots, N-1. \tag{3}$$

The above expression implies that $K_1 \geq K_2 \geq \cdots \geq K_{N-1} \geq 1$. Alternatively, the intermediate buffer capacities $C_n$ can be written in terms of the $K_n$ as follows:

$$C_n = K_n - K_{n+1} \geq 0, n = 1, \ldots, N-2; \quad C_{N-1} = K_{N-1} - 1. \tag{4}$$

Given that the capacity of $E_n$ in the line model is equal to the capacity of echelon buffer $B_n^E$ in the physical line and that the number of tokens in $E_n$ is equal to the echelon WIP following $M_n$ in the physical line, we refer to buffer $E_n, n = 1, \ldots, N-1$, as echelon buffer.

(7) Machine $M_n, n = 2, \ldots, N-1$, is starved if $y_{n-1} = 0$ or, equivalently from (2), if $x_{n-1} = x_n$. Machine $M_N$ is starved if $x_{N-1} = 0$. Machine $M_1$ is never starved and always has one part in it.

(8) Machine $M_n, n = 1, \ldots, N-1$, is blocked before service if $y_{n-1} \geq 1$ and $x_n = K_n$. Machine $M_N$ is never blocked.

Under the above assumptions, $L$ is a discrete-time queueing network with geometrically distributed service times and blocking before service. Each machine $M_n, n = 1, \ldots, N-1$, behaves as a disassembly (split) server because every time $M_n$ produces a part, it also generates a token; the part moves to stage buffer $Q_n$ if machine $M_{n+1}$ is occupied, and the token moves to buffer $E_n$. The vertical line at the end of the system in Figure 3 represents an assembly (merge) operation that assembles parts exiting the network with tokens from all buffers $E_n$. Thus, when a part is produced by machine $M_N$ it draws a token from each of the echelon buffers $E_1, \ldots, E_{N-1}$, signaling that all echelon WIP levels have dropped by one unit. The finished part leaves the network, and the tokens are discarded.

The geometric processing time assumption (3) is the simplest assumption for capturing the randomness of machine processing times. As we will see in Section 6, the method that we develop in this paper for analyzing the system allows us to also deal with the more general case where each machine $M_n$ has load-dependent production probabilities, $p_n(y_{n-1}), n = 2, \ldots, N$. Such a case can be used to model situations where the effective processing times are affected by the workload (Bertrand and Oijen 2002). The existence of such situations has been supported by empirical evidence.

As was mentioned earlier, assumption (6) implies that $K_1 \geq K_2 \geq \cdots \geq K_{N-1} \geq 1$. Now, suppose that $K_n = K_{n+1} = K$ (equivalently, $C_n = 0$, in the physical line) for some $n = 1, \ldots, N-2$. Then, the echelon

WIP levels $x_n$ and $x_{n+1}$ are bounded as follows: $x_n \leq K_n = K$ and $x_{n+1} \leq K_{n+1} = K$. By assumption (8), machine $M_{n+1}$ is blocked if $y_n \geq 1$ and $x_{n+1} = K_{n+1} = K$. If we substitute $y_n$ from (2) and replace $x_{n+1}$ by $K$, then the condition for $M_{n+1}$ to be blocked becomes $x_n - K \geq 1$. However, this condition cannot hold since we assumed that $x_n \leq K$; therefore, if $K_n = K_{n+1} = K$, machine $M_{n+1}$ can never be blocked. In this case, echelon buffer $E_{n+1}$ is obsolete since it never fulfills its role of blocking machine $M_{n+1}$; hence, it can be eliminated. With this in mind, note that the behavior of a network in which $K_n = K \geq 1, n = 1, \ldots, N-1$ (equivalently, $C_n = 0, n = 1, \ldots, N-2$, and $C_{N-1} = C = K - 1 \geq 0$, in the physical line), is equivalent to the behavior of the same network in which all echelon buffers except $E_1$ (equivalently all intermediate buffers except $B_{N-1}$ in the physical line) have been eliminated, as is shown in Figure 2(b). The total WIP following $M_1$ in such a network is capped by $K = 1 + C$; therefore, the physical line operates under a CONWIP policy, as was mentioned in Section 1. Finally, note that although stage buffer $Q_n$ has infinite capacity, the number of parts in it effectively is limited by $K_n$.

To further clarify the connection between the physical production line operated under an EB policy, shown in Figure 1, and the queuing network model of that line, shown in Figure 3, consider the following. Let, $w_{n,m}$ denote the number of parts in intermediate buffer $B_n$ and machine $M_{n+1}$ that have been produced by machine $M_m$ but not by machine $M_{m+1}, m = 1, \ldots, n$, in Figure 1. Also, let $w_n$ denote the total number of parts in buffer $B_n$ and machine $M_{n+1}, n = 1, \ldots, N-1$ (see Figure 1). Then, the following relationships hold:

$$w_n = \sum_{m=1}^{n} w_{n,m}, \quad n = 1, \ldots, N-1, \tag{5}$$

$$y_m = \sum_{n=m}^{N-1} w_{n,m}, \quad m = 1, \ldots, N-1. \tag{6}$$

Finally, from (1) and (6), we have

$$x_n = \sum_{m=n}^{N-1} \sum_{k=m}^{N-1} w_{k,m}, \quad n = 1, \ldots, N-1. \tag{7}$$

Note that the quantities $w_{n,m}$ and $w_n$, like $x_n$ and $y_n$, are functions of the discrete time but we omit this dependence for notational simplicity.

In the following section we develop an approximation method for evaluating the performance of a production line operated under an EB policy based on decomposing the queueing network model described above into easier to solve subsystems.

## 5 Decomposition of the EB-controlled production line model

Let us define the state of the queueing network model of a production line operated under an EB policy described in the previous section as the vector of echelon WIP levels $\mathbf{x} = (x_1, x_2, \ldots, x_{N-1})$. Under assumptions (1)-(8), **x** represents the state of a discrete-time Markov chain. To find the number of states of this chain, we note that the echelon WIP levels satisfy $x_{n+1} \leq x_n \leq K_n, n = 1, \ldots, N-2$, and $0 \leq x_{N-1} \leq K_{N-1}$. From (1), these inequalities can be written in terms of the stage WIP levels $y_n$ as follows: $0 \leq y_n \leq K_n - \sum_{m=n+1}^{N-1} y_m, n = 1, \ldots, N-2$, and $0 \leq y_{N-1} \leq K_{N-1}$. Using these inequalities, we can express the total number of states of the Markov chain under the EB policy, denoted by $NS^E$, as follows:

$$NS^E = \sum_{y_{N-1}=0}^{K_{N-1}} \sum_{y_{N-2}=0}^{K_{N-2}-y_{N-1}} \cdots \sum_{y_n=0}^{K_n-\sum_{m=n+1}^{N-1} y_m} \cdots \sum_{y_2=0}^{K_2-\sum_{m=3}^{N-1} y_m} \sum_{y_1=0}^{K_1-\sum_{m=2}^{N-1} y_m} 1. \tag{8}$$

This number can become very large even for problems of moderate size and is generally significantly larger that the corresponding number of states under the classical IB policy, denoted by $NS^I$, given by

$$NS^I = \prod_{n=1}^{N-1}(C_n + 1). \tag{9}$$

To get an idea of the relative magnitudes of $NS^E$ and $NS^I$, consider a production line with $N = 7$ machines and intermediate buffer capacities $C_n = 5, n = 1, \ldots, 6$, corresponding to echelon buffer capacities $K_1 = 31, K_2 = 26, K_3 = 21, K_4 = 16, K_5 = 11$, and $K_6 = 6$, from (3). From expressions (8) and (9), the number of states for this system under the EB and IB policies is $NS^E = 1{,}404{,}781$ and $NS^I = 46{,}656$, respectively.

Given the explosion in the number of states of the Markov chain model of a production line operated under an EB policy, in this section, we develop an approximation method for evaluating the performance of such a line. This method is based on decomposing the queueing network model of the original line of $N$ machines and $N-1$ echelon buffers into $N-1$ nested segments denoted by $L_n, n = 1, \ldots, N-1$, as shown in Figure 4 for the 4-machine model depicted in Figure 3.

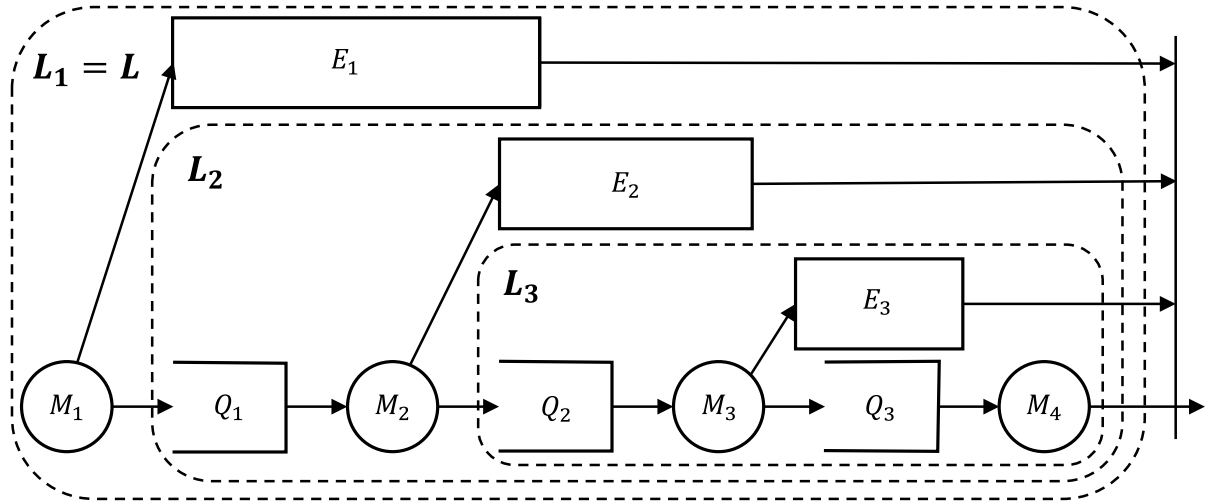

**Figure 4.** Decomposition of the 4-machine production line model shown in Figure 3.

Segment $L_n, n = 2, \ldots, N - 1$, represents the part of the system downstream of machine $M_{n-1}$, while segment $L_1$ represents the entire system. Each segment is then approximated by a two-machine subsystem, denoted by $\tilde{L}_n$, that can be analyzed in isolation.

Figure 5 shows the three subsystems, $\tilde{L}_1, \tilde{L}_2, \tilde{L}_3$, that approximate the three segments, $L_1, L_2, L_3$, shown in Figure 4. Each subsystem can be analyzed independently of the other subsystems, but some of its exogenously defined parameters depend on the analysis of its neighboring subsystems, as we will see later. The ultimate goal of the decomposition is to set the exogenous parameters of each subsystem so that its behavior mimics as closely as possible the behavior of the corresponding segment in the original system.

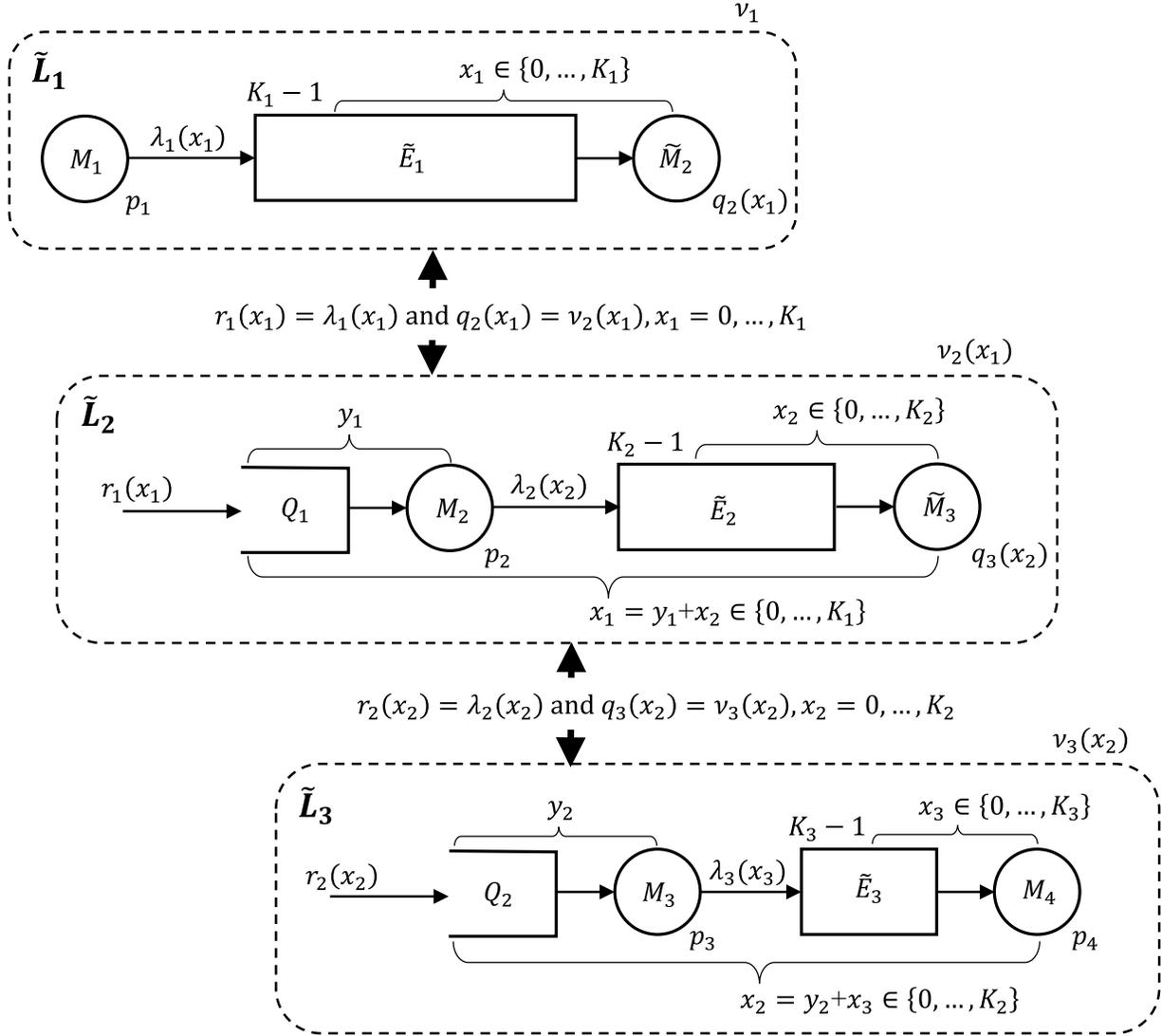

**Figure 5.** Subsystems resulting from the decomposition of the production line model shown in Figure 3.

Let us now take a closer look at the subsystems in Figure 5. Each subsystem $\tilde{L}_n$ has an upstream infinite buffer $Q_{n-1}$ (except for $\tilde{L}_1$ which has none), an upstream machine $M_n$, an intermediate finite buffer $\tilde{E}_n$, and a downstream machine $\widetilde{M}_{n+1}$ (except for the last subsystem $\widetilde{L}_{N-1}$ where the downstream machine is denoted by $M_N$). The first two elements, namely, $Q_{n-1}$ and $M_n$, represent stage buffer $Q_{n-1}$ and machine $M_n$ in segment $L_n$ of the original system; hence, $M_n$ has production probability $p_n$, and the number of parts in $Q_{n-1}$ and $M_n$ is denoted by $y_{n-1}$, where $y_{n-1} = 0, \ldots, K_{n-1}$, as in the original system (see Figure 3).

The ensemble of buffer $\widetilde{E}_n$ and machine $\widetilde{M}_{n+1}$, in subsystem $\widetilde{L}_n, n = 1, \ldots, N-2$, represents in an aggregate way the entire part of the original system downstream of machine $M_n$. This means that it represents segment $L_{n+1}$ which in turn is approximated by subsystem $\widetilde{L}_{n+1}$. It also represents echelon buffer $E_n$ because both $L_{n+1}$ and $E_n$ are fed and depleted simultaneously and hence always have the same number of entities in them. With this in mind, the total capacity of $\widetilde{E}_n$ and $\widetilde{M}_{n+1}$, just like the capacity of $E_n$ and the maximum number of parts in $L_{n+1}$, is $K_n$. More specifically, $\widetilde{E}_n$ has capacity $K_n - 1$ and $\widetilde{M}_{n+1}$ has unit capacity. Moreover, the total number of parts in $\widetilde{E}_n$ and $\widetilde{M}_{n+1}$, just like the number of tokens in $E_n$ and the number of parts in $L_{n+1}$ is denoted by $x_n$, where $x_n = 0, \ldots, K_n$ (see Figure 3).

In the original system, clearly, the higher the value of $x_n$, the more likely it is that a part will come out of segment $L_{n+1}$. To capture this relationship in the approximation method, we assume that $\widetilde{M}_{n+1}$ has a load-dependent production probability denoted by $q_{n+1}(x_n), x_n = 0, \ldots, K_n$. This probability is exogenously defined when analyzing subsystem $\widetilde{L}_n$. As we will see later, eventually, it must be equal to the conditional throughput of subsystem $\widetilde{L}_{n+1}$ (the surrogate of segment $L_{n+1}$ in the original system), which is denoted by $v_{n+1}(x_n)$ (see Figure 5).

In the last subsystem, $\widetilde{L}_{N-1}$, $M_N$ simply represents the last machine in the original system. It is therefore modelled as a simple Bernoulli machine with production probability $p_N$, just like $M_N$ in the original system.

Buffer $Q_{n-1}$ in subsystem $\widetilde{L}_n, n = 2, \ldots, N-1$, receives parts arriving from the outside. The arrival process to this buffer represents in an aggregate way the departure process of parts from machine $M_{n-1}$ in segment $L_{n-1}$ in the original system. An important property of that machine is that it is blocked if echelon buffer $E_{n-1}$ is full, i.e., if $x_{n-1} = K_{n-1}$. To capture this property in the approximation method, we require that the arrival process to buffer $Q_{n-1}$ in subsystem $\widetilde{L}_n, n = 2, \ldots, N-1$, depends on $x_{n-1}$. More specifically, we assume that $Q_{n-1}$ receives parts with a state-dependent arrival probability denoted by $r_{n-1}(x_{n-1}), x_{n-1} = 0, \ldots, K_{n-1}$. This probability is exogenously defined when analyzing $\widetilde{L}_n$, but has the property that $r_{n-1}(K_{n-1}) = 0$. As we will see later, eventually, it must be equal to the internal state-dependent arrival probability of parts to buffer $\widetilde{E}_{n-1}$ in subsystem $\widetilde{L}_{n-1}$ (the surrogate of segment $L_{n-1}$ in the original system) which is denoted by $\lambda_{n-1}(x_{n-1})$ (see Figure 5).

Finally, the total number of parts in subsystem $\widetilde{L}_n, n = 2, \ldots, N-1$, just like the total number of parts in segment $L_n$ of the original system, is denoted by $x_{n-1}$, i.e., $x_{n-1} = y_{n-1} + x_n$, where $x_{n-1} = 0, \ldots, K_{n-1}$, (see also expression (2)).

To evaluate the performance of the original system $L$, we must address the following two problems:

**Problem 1**: How can we analyze each subsystem $\tilde{L}_n$ in isolation given the exogenously defined state-dependent external arrival probabilities $r_{n-1}(x_{n-1}), x_{n-1} = 0, \ldots, K_{n-1}$, (except for $\tilde{L}_1$ that has no external arrivals) and the load-dependent production probabilities of machine $\tilde{M}_{n+1}$, $q_{n+1}(x_n), x_n = 0, \ldots, K_n$ (except for $\tilde{L}_{N-1}$ where machine $M_N$ has production probability $p_N$)?

**Problem 2**: How can we determine the unknown probabilities $r_{n-1}(x_{n-1}), x_{n-1} = 0, \ldots, K_{n-1}, n = 2, \ldots, N-1$, and $q_{n+1}(x_n), x_n = 0, \ldots, K_n, n = 1, \ldots, N-2$?

We address these problems in Sections 6 and 7, respectively. Once these problems have been solved, the performance measures of the original system $L$ can be approximated from the performance measures of subsystems $\tilde{L}_n, n = 1, \ldots, N-1$.

## 6 Analysis of the two-machine subsystems in isolation

In this section, we describe how to analyze each subsystem $\tilde{L}_n, n = 1, \ldots, N-1$, in isolation. First, we concentrate on subsystems $\tilde{L}_n, n = 2, \ldots, N-1$, that have external arrivals, and then we proceed with the simpler subsystem $\tilde{L}_1$ that has no external arrivals.

### 6.1 Analysis of subsystem $\tilde{L}_n, n = 2, \ldots, N-1$

Figure 6 shows the queueing network model of subsystem $\tilde{L}_n$ for the more general case where $M_n$ has load-dependent production probability $p_n(y_{n-1}), n = 2, \ldots, N-1$. We consider this generalization to demonstrate that we can easily apply our analysis to the case where machine $M_n, n = 2, \ldots, N$, in the original model has load-dependent production probability, as was mentioned in Section 4.

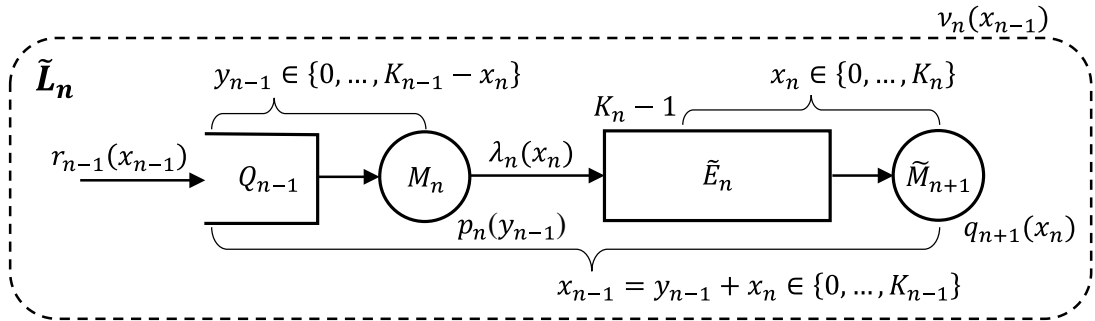

**Figure 6**. Queueing network model of subsystem $\tilde{L}_n, n = 2, \ldots, N-1$.

If we define the state of each subsystem $L^n$ as the vector of the WIP levels $(y_{n-1}, x_n)$, then $(y_{n-1}, x_n)$ represents the state of a two-dimensional discrete-time Markov chain with state-dependent transition probabilities that are functions of the load-dependent production probabilities $p_n(y_{n-1}), y_{n-1} = 0, \ldots, K_{n-1}$, the state-dependent arrival probabilities $r_{n-1}(x_{n-1}), x_{n-1} = 0, \ldots, K_{n-1}$, and the load-dependent production probabilities $q_{n+1}(x_n), x_n = 0, \ldots, K_n$. This Markov chain is irreducible, finite, and aperiodic; therefore unique stationary probabilities exist. The number of states of this chain is $(K_{n-1}+1)(K_n+1) - (K_n+1)K_n/2$. Figure 7 shows the state transition diagram of this chain for $K_{n-1} = 7$ and $K_n = 4$, indicating only the inter-state transitions but not the transition probabilities.

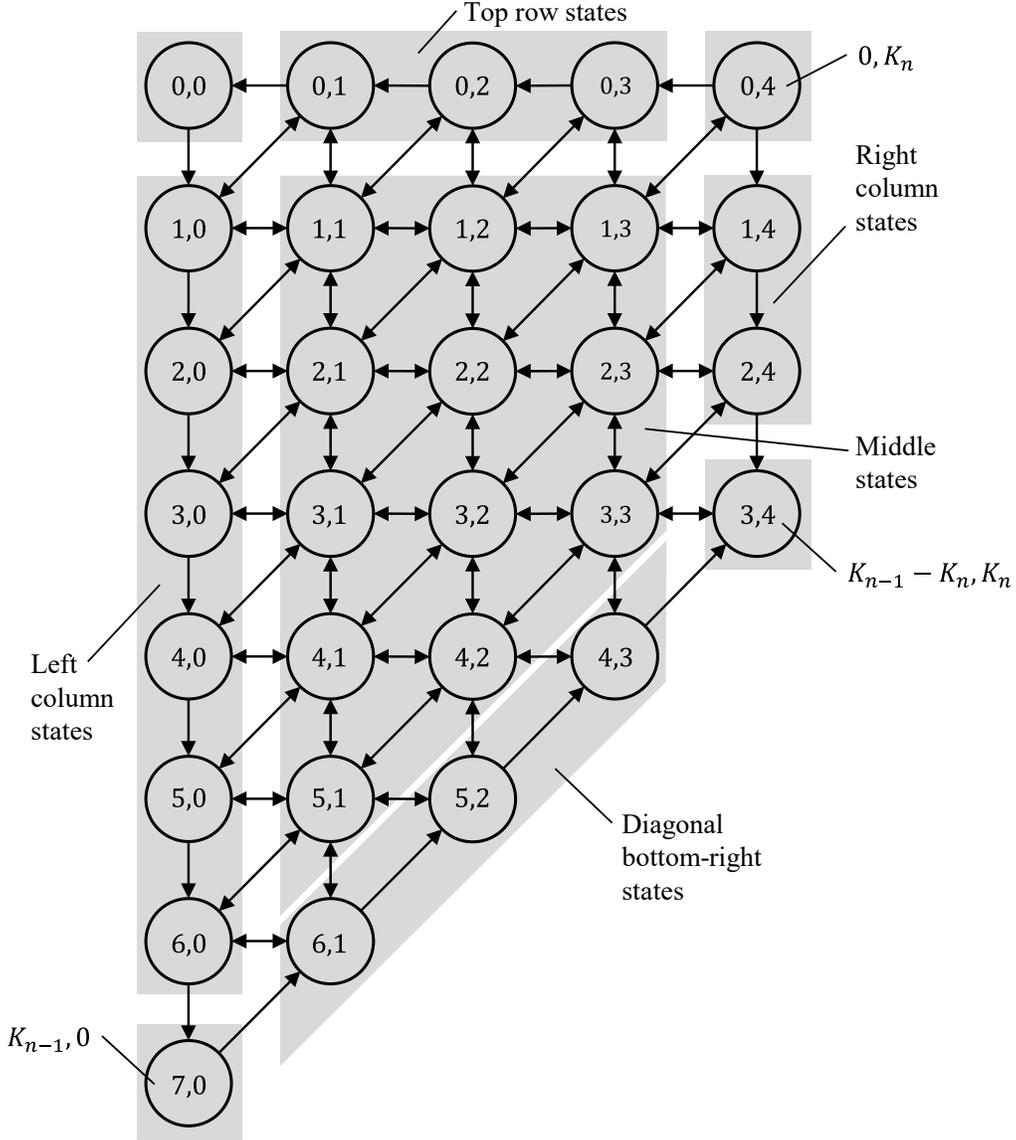

**Figure 7.** State transition diagram of $(y_{n-1}, x_n), n = 2, \ldots, N-1$, for $K_{n-1} = 7$ and $K_n = 4$.

To find the stationary probabilities of this Markov chain, denoted by $P_n(y_{n-1}, x_n)$, we must write the balance equations and the normalization equation and solve them. In what follows, we give the expressions for these equations, where, for notational simplicity, we have:

(1) dropped the subscripts from probabilities $r_{n-1}(\cdot), p_n(\cdot), q_{n+1}(\cdot)$, and $P_n(\cdot, \cdot)$,
(2) used an overbar to indicate the complement of a probability (for instance, $\bar{p} \equiv 1 - p$), and
(3) used $i$ and $j$ to denote states $y_{n-1}$ and $x_n$, respectively.

The form of the balance equations differs depending on whether the states of the Markov chain are in the middle, on the boundaries, or at the corners of the state transition diagram, as is indicated in Figure 7. There are 9 types of states, hence there are 9 types of balance equations. Figures 8-10 show the detailed state transition diagrams for all types of states. Note that the transition probabilities at the extreme states are: $r(K_{n-1}) = p(0) = q(0) = 0$; therefore, $\bar{r}(K_{n-1}) = \bar{p}(0) = \bar{q}(0) = 1$. This fact helps reduce the total number of expressions required to describe all balance equations.

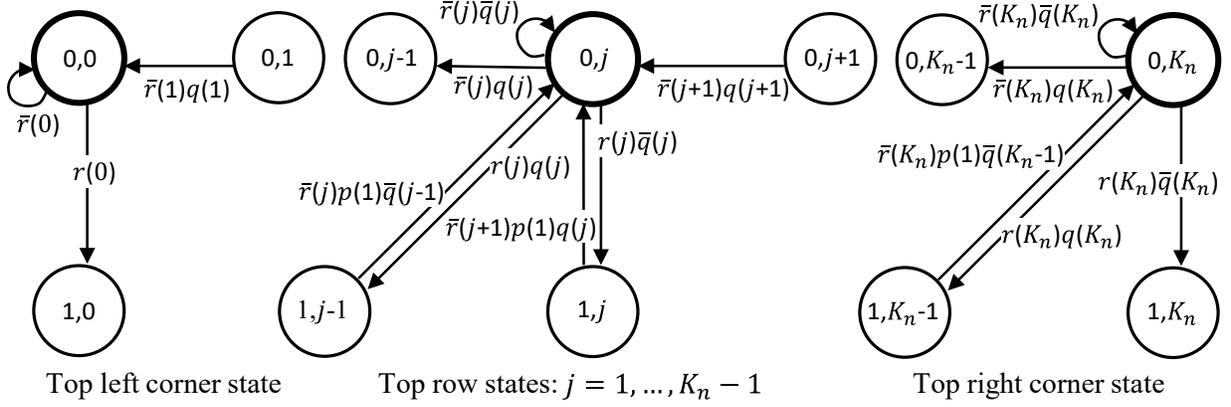

**Figure 8.** State transition diagram of $(i, j)$ for the top states where $i = 0, j = 0, \ldots, K_n$.

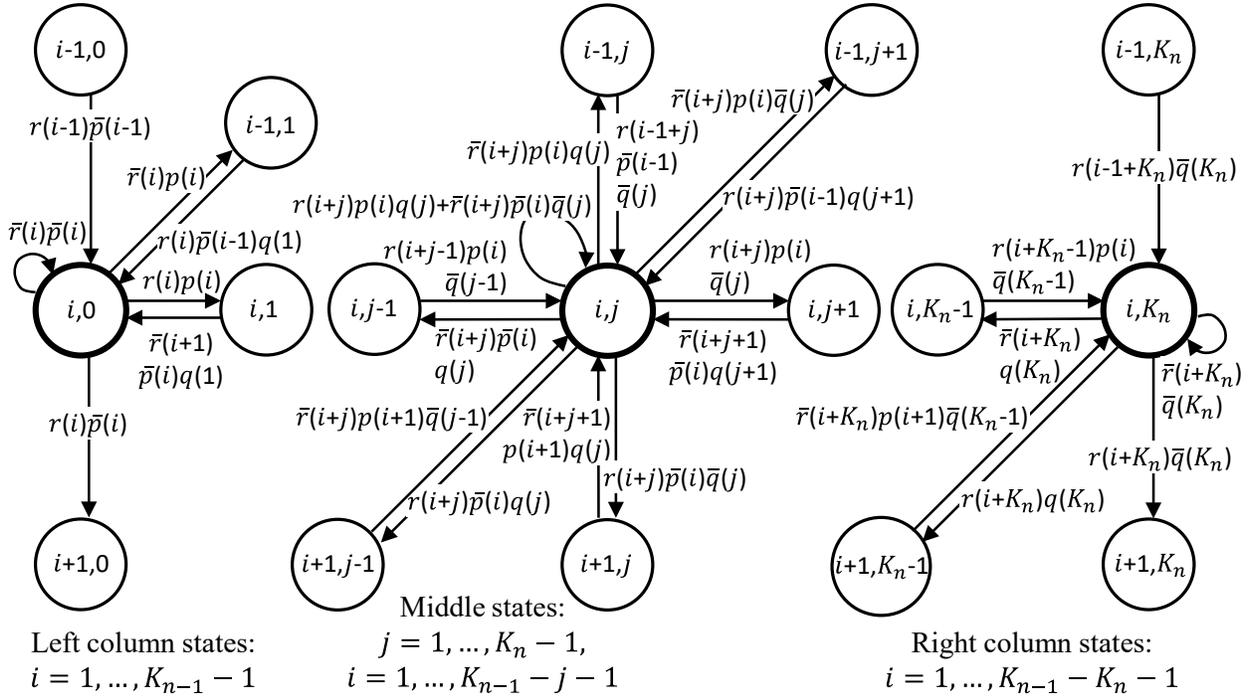

**Figure 9.** State transition diagram of $(i, j)$ for the middle states where $i = 1, \ldots, K_{n-1} - j - 1, j = 0, \ldots, K_n$.

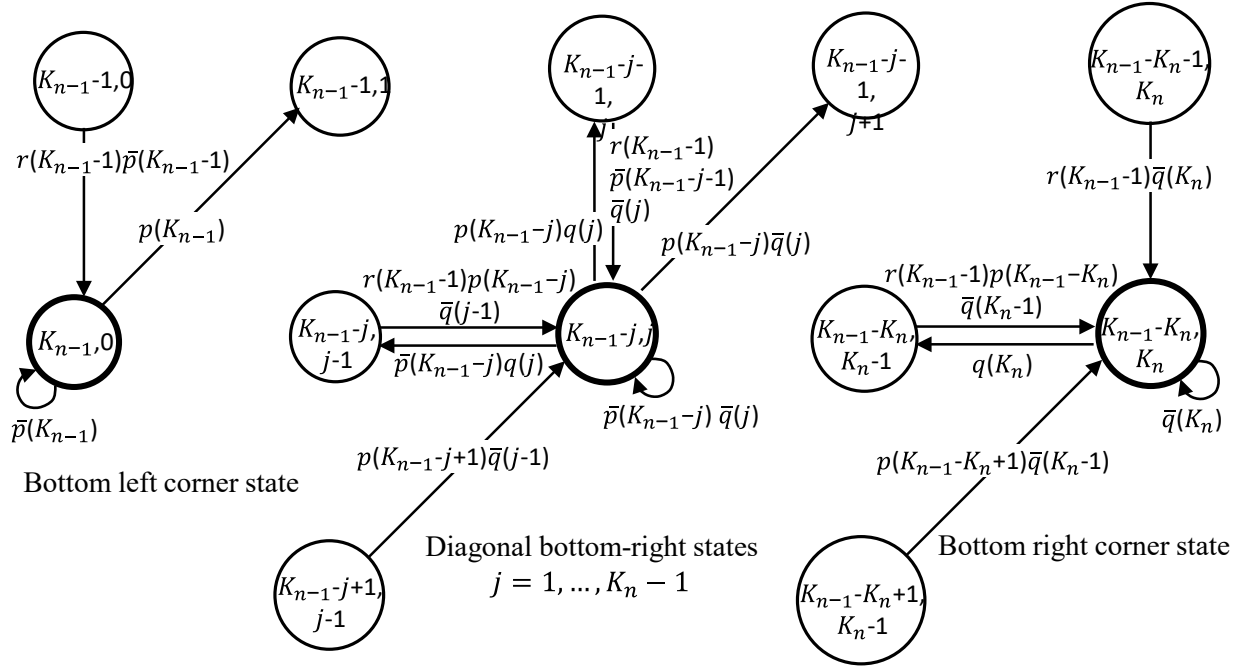

**Figure 10.** State transition diagram of $(i,j)$ for the bottom states where $i = K_{n-1} - j, j = 0, \ldots, K_n$.

The balance equations for the 9 types of states are derived from the state transition diagrams in Figures 8-10 as follows:

**Top left corner state:** $P(0,0)r(0) = P(0,1)\bar{r}(1)q(1)$.

**Top row states:** For $j = 1, \ldots, K_n - 1$,
$$P(0,j)[1 - \bar{r}(j)\bar{q}(j)]$$
$$= P(0, j+1)\bar{r}(j+1)q(j+1) + P(1,j)\bar{r}(j+1)p(1)q(j)$$
$$+ P(1, j-1)\bar{r}(j)p(1)\bar{q}(j-1).$$

**Top right corner state:** $P(0, K_n)[1 - \bar{r}(K_n)\bar{q}(K_n)] = P(1, K_n - 1)\bar{r}(K_n)p(1)\bar{q}(K_n - 1)$.

**Left column states:** For $i = 1, \ldots, K_{n-1} - 1$,
$$P(i, 0)[1 - \bar{r}(i)\bar{p}(i)]$$
$$= P(i-1, 0)r(i-1)\bar{p}(i-1) + P(i-1, 1)r(i)\bar{p}(i-1)q(1)$$
$$+ P(i, 1)\bar{r}(i+1)\bar{p}(i)q(1).$$

**Middle states:** For $j = 1, \ldots, K_n - 1, i = 1, \ldots, K_{n-1} - j - 1$,
$$P(i,j)\big[1 - \big(r(i+j)p(i)q(j) + \bar{r}(i+j)\bar{p}(i)\bar{q}(j)\big)\big]$$
$$= P(i-1, j)r(i+j-1)\bar{p}(i-1)\bar{q}(j) + P(i-1, j+1)r(i+j)\bar{p}(i-1)q(j+1)$$
$$+ P(i, j+1)\bar{r}(i+j+1)\bar{p}(i)q(j+1) + P(i+1, j)\bar{r}(i+j+1)p(i+1)q(j)$$
$$+ P(i+1, j-1)\bar{r}(i+j)p(i+1)\bar{q}(j-1) + P(i, j-1)r(i+j-1)p(i)\bar{q}(j-1)$$

**Right column states:** For $i = 1, \ldots, K_{n-1} - K_n - 1$,

$$P(i, K_n)[1 - \bar{r}(i + K_n)\bar{q}(K_n)]$$
$$= P(i - 1, K_n)r(i + K_n - 1)\bar{q}(K_n) + P(i + 1, K_n - 1)\bar{r}(i + K_n)p(i + 1)\bar{q}(K_n - 1)$$
$$+ P(i, K_n - 1)r(i + K_n - 1)p(i)\bar{q}(K_n - 1).$$

**Bottom left corner state:** $P(K_{n-1}, 0)p(K_{n-1}) = P(K_{n-1} - 1, 0)r(K_{n-1} - 1)\bar{p}(K_{n-1} - 1)$.

**Diagonal bottom-right states:** For $j = 1, \dots, K_n - 1$,
$$P(K_{n-1} - j, j)[1 - \bar{p}(K_{n-1} - j)\bar{q}(j)]$$
$$= P(K_{n-1} - j - 1, j)r(K_{n-1} - 1)\bar{p}(K_{n-1} - j - 1)\bar{q}(j)$$
$$+ P(K_{n-1} - j + 1, j - 1)p(K_{n-1} - j + 1)\bar{q}(j - 1)$$
$$+ P(K_{n-1} - j, j - 1)r(K_{n-1} - 1)p(K_{n-1} - j)\bar{q}(j - 1).$$

**Bottom right corner state:** If $K_{n-1} > K_n$,
$$P(K_{n-1} - K_n, K_n)q(K_n)$$
$$= P(K_{n-1} - K_n - 1, K_n)r(K_{n-1} - 1)\bar{q}(K_n)$$
$$+ P(K_{n-1} - K_n + 1, K_n - 1)p(K_{n-1} - K_n + 1)\bar{q}(K_n - 1)$$
$$+ P(K_{n-1} - K_n, K_n - 1)r(K_{n-1} - 1)p(K_{n-1} - K_n)\bar{q}(K_n - 1).$$

**Normalization equation**
$$\sum_{j=0}^{K_n} \sum_{i=0}^{K_{n-1}-j} P(i, j) = 1.$$

The balance equations above set the steady-state probability flow rate out of each state equal to the steady-state probability flow rate into this state. To see how they are derived, consider the first equation for the top left corner state (0,0) that represents the state where subsystem $\tilde{L}_n$ is totally empty. The only transition out of that state is to state (1,0). This transition occurs if a part arrives in queue $Q_{n-1}$ from the outside. The probability of this event is $r(0)$. Hence, the probability flow rate out of state (0,0) is $P(0,0)r(0)$. The only transition into state (0,0) is from state (0,1). This transition occurs if no part arrives in $Q_{n-1}$ and machine $\tilde{M}_{n+1}$ produces the part in it. The probability of this event is $\bar{r}(1)p(1)$. Hence the probability flow rate into state (0,0) is $P(0,1)\bar{r}(1)q(1)$. The balance equations is therefore $P(0,0)r(0) = P(0,1)\bar{r}(1)q(1)$. The other equations are derived similarly by taking into account the three types of events that may or may not occur in each period, namely, the arrival of a part in $Q_{n-1}$, the production of a part by $M_n$, and the production of a part $\tilde{M}_{n+1}$.

The above system of equations is linear and has a unique solution. It can be solved using any numerical analysis scheme. In our numerical examples, we use the Gauss-Seidel method, where in each iteration we sequentially update the stationary probability of each state using the most recent values of the stationary probabilities of the other states involved. At the end of each iteration, we normalize all probabilities. We terminate the iterations when the maximum absolute percentage difference between two successive iterations is below a very small number $\varepsilon$. Once we have computed the stationary probabilities, we can use them to calculate the following performance measures of interest:

$v_n(x_{n-1}), x_{n-1} = 0, \ldots, K_{n-1}$: conditional throughput of subsystem $\tilde{L}_n$.

$\lambda_n(x_n), x_n = 0, \ldots, K_n$: internal state-dependent arrival probability of parts to buffer $\tilde{E}_n$.

$\bar{x}_n$: average WIP level of buffer $\tilde{E}_n$.

$\theta_{n-1}$: overflow probability of buffer $Q_{n-1}$ defined as the probability that $y_{n-1}$ will increase by one unit when $y_{n-1} \geq K_{n-1} - K_n + 1 = C_{n-1} + 1$; $\theta_{n-1}$ represents the probability that a part will be produced by machine $M_{n-1}$ and will be physically transferred for storage in an intermediate buffer downstream of $B_{n-1}$ because $B_{n-1}$ is full (hence, the term "overflow"). This probability is important especially if the transportation cost associated with this transfer is significant.

Note that in the above definitions, we have restored the original notation. The above performance measures are computed as follows:

$$\lambda_n(x_n) = \begin{cases} \dfrac{\sum_{y_{n-1}=0}^{K_{n-1}-x_n} P_n(y_{n-1}, x_n) p_n(y_{n-1})}{\sum_{y_{n-1}=0}^{K_{n-1}-x_n} P_n(y_{n-1}, x_n)}, & x_n = 0, \ldots, K_n - 1, \\ 0, & x_n = K_n, \end{cases} \quad (10)$$

$$v_n(x_{n-1}) = \dfrac{\sum_{y_{n-1}=(x_{n-1}-K_n)^+}^{x_{n-1}} P_n(y_{n-1}, x_{n-1} - y_{n-1}) q_{n+1}(x_{n-1} - y_{n-1})}{\sum_{y_{n-1}=(x_{n-1}-K_n)^+}^{x_{n-1}} P_n(y_{n-1}, x_{n-1} - y_{n-1})}, \quad (11)$$

$$x_{n-1} = 0, \ldots, K_{n-1},$$

$$\bar{x}_n = \sum_{x_n=0}^{K_n} x_n \sum_{y_{n-1}=0}^{K_{n-1}-x_n} P_n(y_{n-1}, x_n), \quad (12)$$

$$\theta_{n-1} = \sum_{x_n=0}^{K_n} \sum_{y_{n-1}=K_{n-1}-K_n+1}^{K_{n-1}-x_n} P_n(y_{n-1}, x_n) r_{n-1}(y_{n-1} + x_n) \bar{p}_n(y_{n-1}). \quad (13)$$

The derivations of the above expressions can be found in Section S1 of the online supplement.

## 6.2 Analysis of subsystem $\tilde{L}_1$

The first subsystem of the decomposition, $\tilde{L}_1$, shown at the top of Figure 5, differs from the other subsystems in that there is no input process to machine $M_1$; hence, it is simpler. Just like $M_1$ in the original system, machine $M_1$ in subsystem $\tilde{L}_1$ is never starved and in every period produces a part with probability $p_1$ unless it is blocked when buffer $\tilde{E}_1$ is full. If we define the state of $\tilde{L}_1$ as the WIP level $x_1$, then $x_1$ represents the state of a discrete-time finite-state birth-death process, for which the stationary probabilities, denoted by $P_1(x_1)$, can be easily computed. The state transition diagram of this Markov chain is shown in

Figure 11. As previously, for notational simplicity, we dropped the subscripts from probabilities $p_1$, $q_2(\cdot)$, and $P_1(\cdot)$. We also used an overbar to indicate the complement of a probability, and $j$ to denote state $x_1$.

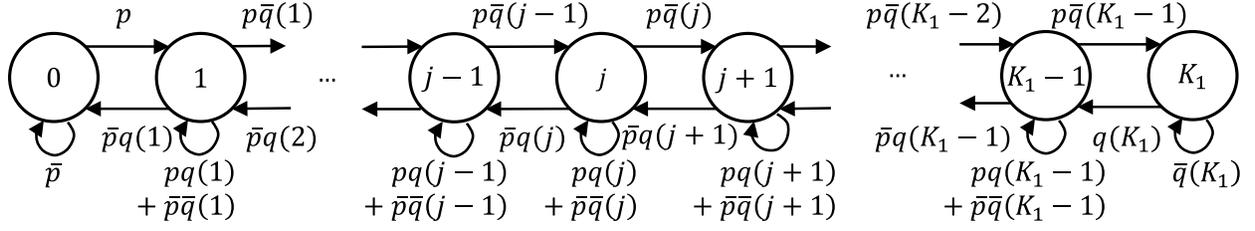

**Figure 11.** State transition diagram of $x_1$.

To compute the stationary probabilities of $x_1$, we define the following coefficients:

$$G(j) = \begin{cases} \dfrac{p^j \prod_{i=0}^{j-1} \bar{q}(i)}{\bar{p}^j \prod_{i=1}^{j} q(i)}, & j = 0, \ldots, K_1 - 1, \\[1em] \dfrac{p^{K_1} \prod_{i=0}^{K_1-1} \bar{q}(i)}{\bar{p}^{K_1-1} \prod_{i=1}^{K_1} q(i)}, & j = K_1. \end{cases}$$

In the above expressions, we have exploited the fact that $q(0) = 0$; therefore, $\bar{q}(0) = 1$. The stationary probabilities are given by

$$P(j) = \frac{G(j)}{\sum_{i=0}^{K_1} G(i)}, \quad j = 0, \ldots, K_1.$$

Once we have computed the stationary probabilities, we can use them to calculate the average throughput of subsystem $\tilde{L}_1$, denoted by $\nu_1$, and the average WIP level in echelon buffer $\tilde{E}_1$, denoted by $\bar{x}_1$, where we have restored the original notation. These two measures are calculated as follows:

$$\nu_1 = p_1\big(1 - P_1(K_1)\big), \tag{14}$$

$$\bar{x}_1 = \sum_{x_1=0}^{K_1} x_1 P_1(x_1). \tag{15}$$

Finally, note that the internal state-dependent arrival probability of parts to buffer $\tilde{E}_1$, denoted by $\lambda_1(x_1), x_1 = 0, \ldots, K_1$, is simply given by

$$\lambda_1(x_1) = \begin{cases} p_1, & x_1 = 0, \ldots, K_1 - 1, \\ 0, & x_1 = K_1. \end{cases} \tag{16}$$

## 7  Analysis of the entire original EB-controlled production line model

The unknown parameters of each subsystem $\tilde{L}_n$ are the state-dependent external arrival probabilities $r_{n-1}(x_{n-1}), x_{n-1} = 0, \ldots, K_{n-1}$ (except in $\tilde{L}_1$, where there are no external arrivals), and the load-dependent production rates $q_{n+1}(x_n), x_n = 0, \ldots, K_n$, of the downstream machine $\tilde{M}_{n+1}$ (except in $\tilde{L}_{N-1}$, where the downstream machine is identical to machine $M_N$ in the original system, and therefore has known production probability $p_N$). To determine the values of these parameters we set up a system of equations that relate the flow of parts in subsystem $\tilde{L}_n$ with the flow of parts in the neighboring subsystems $\tilde{L}_{n+1}$ and $\tilde{L}_{n-1}$.

More specifically, as we wrote earlier, in subsystem $\tilde{L}_n, n = 1, \ldots, N-2$, $\tilde{M}_{n+1}$ is an aggregate representation of subsystem $\tilde{L}_{n+1}$, which is the surrogate of segment $L_{n+1}$ in the original system. The load-dependent production probabilities of $\tilde{M}_{n+1}$, $q_{n+1}(x_n), x_n = 0, \ldots, K_n$, should therefore be equal to the conditional throughput of $\tilde{L}_{n+1}$, $v_{n+1}(x_n), x_n = 0, \ldots, K_n$. Similarly, in subsystem $\tilde{L}_{n-1}, n = 2, \ldots, N-1$, $\tilde{M}_n$ is an aggregate representation of subsystem $\tilde{L}_n$. The external arrival process to buffer $B_{n-1}$ in $\tilde{L}_n$, $r_{n-1}(x_{n-1}), x_{n-1} = 0, \ldots, K_{n-1}$, should therefore be equal to the internal state-dependent arrival process of parts from machine $M_{n-1}$ to buffer $\tilde{E}_{n-1}$ in $\tilde{L}_{n-1}$, $\lambda_{n-1}(x_{n-1}), x_{n-1} = 0, \ldots, K_{n-1}$.

The above relationships can be written as follows:

$$q_{n+1}(x_n) = v_{n+1}(x_n), \quad x_n = 0, \ldots, K_n, \quad n = 1, \ldots, N-2, \tag{17}$$

$$r_{n-1}(x_{n-1}) = \lambda_{n-1}(x_{n-1}), \quad x_{n-1} = 0, \ldots, K_{n-1}, \quad n = 2, \ldots, N-1. \tag{18}$$

For each subsystem $\tilde{L}_n$, the conditional throughput $v_n(x_{n-1})$ and the internal state-dependent arrival probability $\lambda_n(x_n)$ can be computed by analyzing the subsystem in isolation, given the values of the production probabilities $p_n(y_{n-1}), y_{n-1} = 0, \ldots, K_{n-1}$, and $q_{n+1}(x_n), x_n = 0, \ldots, K_n$, as was shown in Section 6. This means that $v_{n+1}(x_n)$ in (17) is a function of $p_{n+1}(y_n)$ and $q_{n+2}(x_{n+1})$. Also, $\lambda_{n-1}(x_{n-1})$ in (18) is a function of $p_{n-1}(y_{n-2})$ and $q_n(x_{n-1})$. Hence, the unknown parameters $r_{n-1}(x_{n-1})$ and $q_{n+1}(x_n)$ in expressions (17) and (18) are the solution of a fixed-point problem. To determine their values, we use the following iterative algorithm.

**Algorithm for analyzing the entire production system**

**Step 1.** Initialization:

1.1. Set the unknown external arrival probabilities of each subsystem $\tilde{L}_n$ (except $\tilde{L}_1$ which receives no external arrivals) to some initial value. A reasonable initial value that we have used in our numerical experiments is the smallest production probability of all machines upstream of $M_n$, namely,

$$r_{n-1}^{\text{init}}(x_{n-1}) = \begin{cases} \min\{p_m : m = 1, \ldots, n-1\}, & x_{n-1} = 0, \ldots, K_{n-1} - 1, \\ 0, & x_{n-1} = K_{n-1}, \end{cases} \quad n = 2, \ldots, N-1. \tag{19}$$

1.2. Set the unknown production rates of machine $\tilde{M}_{n+1}$ in each subsystem $\tilde{L}_1$ (except $\tilde{L}_{N-1}$ where the production rate of the downstream machine is $p_N$) to some initial value. A reasonable initial value that we have used in our experiments is the smallest production probability of all machines downstream of $M_n$, namely,

$$q_{n+1}^{\text{init}}(x_n) = \begin{cases} 0, & x_n = 0, \\ \min\{p_m : m = n+1, \ldots, N\}, & x_n = 1, \ldots, K_n, \end{cases} \quad n = 1, \ldots, N-2. \tag{20}$$

**Step 2.** Main Iteration: Iterate backwards and forwards until the external and internal arrival probabilities converge, i.e., until $r_{n-1}(x_{n-1}) = \lambda_{n-1}(x_{n-1}), x_{n-1} = 0, \ldots, K_{n-1}, n = 2, \ldots, N-1$. More specifically,

**Set** $n = N - 1$.

**While** $n \geq 2$,

    **If** $n = N - 1$,

        Given $r_{N-2}(x_{N-2}), x_{N-2} = 0, \ldots, K_{N-2}$, **solve** subsystem $\tilde{L}_{N-1}$ and **compute** $v_{N-1}(x_{N-2}), x_{N-2} = 1, \ldots, K_{N-2}, \bar{x}_{N-1}$, and $\theta_{N-1}$ from (11)-(13), respectively, for $n = N - 1$.

        **Set** $q_{N-1}(x_{N-2}) = v_{N-1}(x_{N-2}), x_{N-2} = 1, \ldots, K_{N-2}$; **set** $n = N - 2$.

    **Else**

        Given $r_{n-1}(x_{n-1}), x_{n-1} = 0, \ldots, K_{n-1}$, and $q_{n+1}(x_n), x_n = 0, \ldots, K_n$, **solve** subsystem $\tilde{L}_n$ and **compute** $\lambda_n(x_n), x_n = 0, \ldots, K_n - 1, v_n(x_{n-1}), x_{n-1} = 1, \ldots, K_{n-1}, \bar{x}_n$, and $\theta_n$ from (10)-(13), respectively.

        **If** $\lambda_n(x_n) \approx r_n(x_n), x_n = 0, \ldots, K_n - 1$,

            **Set** $q_n(x_{n-1}) = v_n(x_{n-1}), x_{n-1} = 0, \ldots, K_{n-1}$; **set** $n = n - 1$.

        **Else**

            **Set** $r_n(x_n) = \lambda_n(x_n), x_n = 0, \ldots, K_n$; **set** $n = n + 1$.

        **Endif**

    **Endif**

**Endwhile**

**Step 3.** Compute average system throughput and WIP: Given $q_2(x_1), x_1 = 0, \ldots, K_1$, solve subsystem $\tilde{L}_1$ and compute average throughput $v_1$ from (14) and average WIP level $\bar{x}_1$ from (15). These two values are the estimates of the average throughput and total WIP of the system. Similarly, the final values of $\bar{x}_n, n = 2, \ldots, N - 1$, are the estimates of the average echelon WIP downstream of machine $M_n, n = 2, \ldots, N - 1$. From (1) and (2), we can also obtain estimates of the average stage WIP levels denoted by $\bar{y}_n, n = 1, \ldots, N -$ 1, as follows: $\bar{y}_n = \bar{x}_n - \bar{x}_{n+1}, n = 1, \ldots, N-1$, and $\bar{y}_{N-1} = \bar{x}_{N-1}$. Finally, the final values of $\theta_n, n = 2, \ldots, N-1$, are the estimates of the overflow probabilities of $B_{n-1}, n = 2, \ldots, N-1$.

Note that the first time each subsystem $\tilde{L}_n, n = 2, \ldots, N-1$, is solved using the method presented in Section 6, the stationary probabilities of the Markov chain whose states are $(y_{n-1}, x_n)$ must be initialized. The simplest way to do this is to set them all equal so that their sum is one. A more sophisticated way is to set $P_n(y_{n-1}, x_n)$ equal to the normalized product of the approximate marginal stationary distributions of $y_{n-1}$ and $x_n$ in isolation. The approximate marginal distribution of $y_{n-1}$ in isolation can be found by solving a two-machine one-buffer line (as a discrete-time finite-state birth-death process), where the upstream and downstream machines have production probabilities $r_{n-1}^{\text{init}}(x_{n-1})$ and $q_n^{\text{init}}(x_{n-1})$ given by (19) and (20), respectively. Similarly, the approximate marginal distribution of $x_n$ in isolation can be found by solving a two-machine one-buffer line, where the upstream and downstream machines have production probabilities $r_n^{\text{init}}(x_n)$ and $q_{n+1}^{\text{init}}(x_n)$ given by (19) and (20), respectively. These problems can be solved extremely fast. From then on, each time subsystem $\tilde{L}_n, n = 2, \ldots, N-1$, is solved again, the stationary probabilities from the previous time are used as initial values. Numerical experimentation has shown that this method results in significant gains in overall computational time.

Finally, the criterion that we used to detect if $\lambda_n(x_n) \approx r_n(x_n), x_n = 0, \ldots, K_n - 1$, in step 2 of the above procedure is $\max_{x_n = 0, \ldots, K_n - 1}\{|\lambda_n(x_n) - r_n(x_n)|/r_n(x_n)\} < \varepsilon$, where $\varepsilon$ is a very small number.

## 8 Numerical results

In this section, we evaluate the accuracy and efficiency of the decomposition method developed in Sections 6 and 7 by comparing it against simulation, for several instances of two numerical examples, also exploring the effect of system parameters on system performance. In all instances, we used the value of $\varepsilon = 0.0001$ for the convergence criterion both in the procedure for analyzing each subsystem $\tilde{L}_n$ in isolation, described in Section 6.1, and in the algorithm for analyzing the original system $L$, described in Section 7. Regarding the convergence of these two algorithms, we know from Markov chain theory that the balance equations have a unique solution because the underlying Markov chain is irreducible, finite and aperiodic. This means that the procedure for analyzing each subsystem $\tilde{L}_n$ in isolation should always converge. Although we cannot similarly guarantee the convergence of the algorithm for analyzing the entire system, we can attest that in all the instances that we run, the algorithm converged. Both the decomposition and simulation algorithms were written in Matlab R2012b and were run on a PC with an Intel(R) Core(TM) i7-920 CPU @ 2.67 GHz.

In Example 1, we consider a production line consisting of $N = 5$ machines and 4 buffers. For this system, we evaluated 34 different instances (cases). Table 1 shows the input data for each case, namely, the production probabilities of the machines, $p_n, n = 1, \ldots, 5$, the capacities of the intermediate buffers, $C_n, n = 1, \ldots, 4$, and the resulting capacities of echelon buffers, $K_n, n = 1, \ldots, 4$, computed from (3).

Table 1. Input data for the 5-machine line Example 1.

| # | $p_1$ | $p_2$ | $p_3$ | $p_4$ | $p_5$ | $C_1$ | $C_2$ | $C_3$ | $C_4$ | $K_1$ | $K_2$ | $K_3$ | $K_4$ |
|---|---|---|---|---|---|---|---|---|---|---|---|---|---|
| 1 | 0.6 | 0.6 | 0.6 | 0.6 | 0.6 | 1 | 1 | 1 | 1 | 5 | 4 | 3 | 2 |
| 2 | 0.6 | 0.6 | 0.6 | 0.6 | 0.6 | 5 | 5 | 5 | 5 | 21 | 16 | 11 | 6 |
| 3 | 0.6 | 0.6 | 0.6 | 0.6 | 0.6 | 10 | 10 | 10 | 10 | 41 | 31 | 21 | 11 |
| 4 | 0.6 | 0.6 | 0.6 | 0.6 | 0.6 | 15 | 15 | 15 | 15 | 61 | 46 | 31 | 16 |
| 5 | **0.4** | 0.6 | 0.6 | 0.6 | 0.6 | 1 | 1 | 1 | 1 | 5 | 4 | 3 | 2 |
| 6 | **0.4** | 0.6 | 0.6 | 0.6 | 0.6 | 5 | 5 | 5 | 5 | 21 | 16 | 11 | 6 |
| 7 | **0.4** | 0.6 | 0.6 | 0.6 | 0.6 | 10 | 10 | 10 | 10 | 41 | 31 | 21 | 11 |
| 8 | 0.6 | 0.6 | **0.4** | 0.6 | 0.6 | 1 | 1 | 1 | 1 | 5 | 4 | 3 | 2 |
| 9 | 0.6 | 0.6 | **0.4** | 0.6 | 0.6 | 5 | 5 | 5 | 5 | 21 | 16 | 11 | 6 |
| 10 | 0.6 | 0.6 | **0.4** | 0.6 | 0.6 | 10 | 10 | 10 | 10 | 41 | 31 | 21 | 11 |
| 11 | 0.6 | 0.6 | 0.6 | 0.6 | **0.4** | 1 | 1 | 1 | 1 | 5 | 4 | 3 | 2 |
| 12 | 0.6 | 0.6 | 0.6 | 0.6 | **0.4** | 5 | 5 | 5 | 5 | 21 | 16 | 11 | 6 |
| 13 | 0.6 | 0.6 | 0.6 | 0.6 | **0.4** | 10 | 10 | 10 | 10 | 41 | 31 | 21 | 11 |
| 14 | **0.4** | 0.5 | 0.6 | 0.7 | 0.8 | 5 | 5 | 5 | 5 | 21 | 16 | 11 | 6 |
| 15 | 0.8 | 0.7 | 0.6 | 0.5 | **0.4** | 5 | 5 | 5 | 5 | 21 | 16 | 11 | 6 |
| 16 | 0.8 | 0.8 | 0.8 | 0.8 | 0.8 | 10 | 10 | 10 | 10 | 41 | 31 | 21 | 11 |
| 17 | 0.6 | 0.6 | 0.6 | 0.6 | 0.6 | 0 | 0 | 0 | 4 | 5 | 5 | 5 | 5 |
| 18 | **0.4** | 0.6 | 0.6 | 0.6 | 0.6 | 0 | 0 | 0 | 4 | 5 | 5 | 5 | 5 |
| 19 | 0.6 | 0.6 | **0.4** | 0.6 | 0.6 | 0 | 0 | 0 | 4 | 5 | 5 | 5 | 5 |
| 20 | 0.6 | 0.6 | 0.6 | 0.6 | **0.4** | 0 | 0 | 0 | 4 | 5 | 5 | 5 | 5 |
| 21 | **0.4** | 0.5 | 0.6 | 0.7 | 0.8 | 0 | 0 | 0 | 4 | 5 | 5 | 5 | 5 |
| 22 | 0.8 | 0.7 | 0.6 | 0.5 | **0.4** | 0 | 0 | 0 | 4 | 5 | 5 | 5 | 5 |
| 23 | 0.6 | 0.6 | 0.6 | 0.6 | 0.6 | 0 | 0 | 0 | 20 | 21 | 21 | 21 | 21 |
| 24 | **0.4** | 0.6 | 0.6 | 0.6 | 0.6 | 0 | 0 | 0 | 20 | 21 | 21 | 21 | 21 |
| 25 | 0.6 | 0.6 | **0.4** | 0.6 | 0.6 | 0 | 0 | 0 | 20 | 21 | 21 | 21 | 21 |
| 26 | 0.6 | 0.6 | 0.6 | 0.6 | **0.4** | 0 | 0 | 0 | 20 | 21 | 21 | 21 | 21 |
| 27 | **0.4** | 0.5 | 0.6 | 0.7 | 0.8 | 0 | 0 | 0 | 20 | 21 | 21 | 21 | 21 |
| 28 | 0.8 | 0.7 | 0.6 | 0.5 | **0.4** | 0 | 0 | 0 | 20 | 21 | 21 | 21 | 21 |
| 29 | 0.6 | 0.6 | 0.6 | 0.6 | 0.6 | 0 | 0 | 0 | 40 | 41 | 41 | 41 | 41 |
| 30 | **0.4** | 0.6 | 0.6 | 0.6 | 0.6 | 0 | 0 | 0 | 40 | 41 | 41 | 41 | 41 |
| 31 | 0.6 | 0.6 | **0.4** | 0.6 | 0.6 | 0 | 0 | 0 | 40 | 41 | 41 | 41 | 41 |
| 32 | 0.6 | 0.6 | 0.6 | 0.6 | **0.4** | 0 | 0 | 0 | 40 | 41 | 41 | 41 | 41 |
| 33 | **0.4** | 0.5 | 0.6 | 0.7 | 0.8 | 0 | 0 | 0 | 40 | 41 | 41 | 41 | 41 |
| 34 | 0.8 | 0.7 | 0.6 | 0.5 | **0.4** | 0 | 0 | 0 | 40 | 41 | 41 | 41 | 41 |

The cases are divided into three groups as far as the distribution of the production probabilities among the machines is concerned. Cases 1-4, 16-17, 23, and 29 represent balanced lines where all machines have the same production probabilities. Cases 5-13, 18-20, 24-26, and 30-32 represent lines where all the machines have the same production probabilities, except for one that has a smaller probability, making it the slower machine. That machine is either the first, the middle, or the last. Finally, cases, 14-15, 21-22, 27-28, and 33-34 represent unbalanced lines where the machines have increasing or decreasing production probabilities.

In terms of the intermediate buffer capacity allocation, the cases are divided into two groups. Cases 1-16 represent lines where the capacities of all intermediate buffers are the same, implying that the echelon buffer capacities increase by the same amount as we move upstream the line. In cases 17-34, the capacities of all intermediate buffers except the last one are zero, implying that all echelon buffer capacities are the same. As was mentioned in Section 4, this corresponds to a line operating under CONWIP.

Table 2 shows the performance measure estimates of the EB policy obtained by decomposition. These measures are the average stage WIP levels, denoted by $\bar{y}_n, n = 1, \ldots, 4$, the average line throughput, denoted by $v$, the average overflow rate of buffer $B_n$, denoted by $\theta_n, n = 1, \ldots, 3$, and the computation time, CPU, in seconds. We report the average stage WIP levels rather than the average total WIP levels because often the inventory holding cost rate differs at different stages; typically, it is increasing in the stages because of the value added at each stage. Therefore, it is important to explore the accuracy of the decomposition method at the individual stage level rather than at the level of the entire production line. Recall that the values of $\bar{y}_n$, $v$, and $\theta_n$ are computed as the final values of $\bar{y}_n$, $v_1$, and $\theta_n$ in the algorithm described in Section 7. Note that $\theta_4$ in the 5-machine example, and more generally $\theta_{N-1}$ in the $N$-machine case, is zero because there is no overflow of parts in the last buffer $B_{N-1}$.

**Table 2.** Performance measure estimates of the EB policy for the 5-machine line Example 1 obtained by decomposition.

| # | $\bar{y}_1$ | $\bar{y}_2$ | $\bar{y}_3$ | $\bar{y}_4$ | $v$ | $\theta_1$ | $\theta_2$ | $\theta_3$ | CPU (s) |
|---|---|---|---|---|---|---|---|---|---|
| 1 | 1.23163 | 1.09870 | 0.97433 | 0.77488 | 0.38037 | 0.03426 | 0.02731 | 0.01882 | 0.004 |
| 2 | 5.48583 | 4.89215 | 4.25719 | 2.65946 | 0.54193 | 0.08878 | 0.07539 | 0.06011 | 0.045 |
| 3 | 10.82805 | 9.68193 | 8.42532 | 4.95217 | 0.56994 | 0.10380 | 0.08936 | 0.07294 | 0.380 |
| 4 | 16.13622 | 14.49405 | 12.61259 | 7.24265 | 0.57974 | 0.10976 | 0.09496 | 0.07826 | 1.368 |
| 5 | 0.89084 | 0.85861 | 0.80076 | 0.66638 | 0.33398 | 0.01941 | 0.01716 | 0.01263 | 0.003 |
| 6 | 1.20063 | 1.20419 | 1.22340 | 1.17593 | 0.39999 | 0.00185 | 0.00187 | 0.00194 | 0.021 |
| 7 | 1.20000 | 1.20000 | 1.20049 | 1.19952 | 0.40000 | 0.00003 | 0.00003 | 0.00003 | 0.076 |
| 8 | 1.14907 | 1.73232 | 0.75492 | 0.64002 | 0.32303 | 0.02261 | 0.07934 | 0.01077 | 0.003 |
| 9 | 5.01199 | 12.40126 | 1.21637 | 1.17314 | 0.39974 | 0.03620 | 0.23718 | 0.00179 | 0.021 |
| 10 | 10.00018 | 27.39982 | 1.20048 | 1.19952 | 0.40000 | 0.03622 | 0.24000 | 0.00003 | 0.091 |
| 11 | 1.14272 | 1.05090 | 0.97693 | 1.12223 | 0.31788 | 0.02169 | 0.01755 | 0.01277 | 0.004 |
| 12 | 5.01007 | 4.99346 | 4.97442 | 4.83558 | 0.39836 | 0.03579 | 0.03530 | 0.03476 | 0.023 |
| 13 | 10.00012 | 9.99980 | 9.99902 | 9.80129 | 0.39997 | 0.03621 | 0.03621 | 0.03620 | 0.092 |
| 14 | 2.38749 | 1.19888 | 0.79971 | 0.59974 | 0.39992 | 0.02089 | 0.00183 | 0.00013 | 0.013 |
| 15 | 5.18497 | 5.34579 | 5.73138 | 4.15700 | 0.39074 | 0.02334 | 0.04523 | 0.08401 | 0.013 |
| 16 | 10.79251 | 9.65960 | 8.40536 | 4.98270 | 0.77899 | 0.07185 | 0.06185 | 0.05039 | 0.401 |
| 17 | 1.00000 | 1.00001 | 1.00001 | 0.99999 | 0.40050 | 0.10074 | 0.10074 | 0.10074 | 0.004 |
| 18 | 0.78465 | 0.78465 | 0.78464 | 0.78464 | 0.34079 | 0.07323 | 0.07323 | 0.07323 | 0.003 |
| 19 | 0.78464 | 1.86144 | 0.78464 | 0.78464 | 0.34079 | 0.07323 | 0.16477 | 0.07323 | 0.003 |
| 20 | 0.78464 | 0.78466 | 0.78462 | 1.86143 | 0.34079 | 0.07323 | 0.07323 | 0.07323 | 0.004 |
| 21 | 1.16030 | 0.80390 | 0.60786 | 0.48619 | 0.34526 | 0.11341 | 0.07561 | 0.04860 | 0.003 |
| 22 | 0.60787 | 0.80387 | 1.16034 | 1.94171 | 0.34525 | 0.04860 | 0.07561 | 0.11341 | 0.004 |
| 23 | 4.19933 | 4.20121 | 4.19807 | 4.20086 | 0.55281 | 0.20335 | 0.20336 | 0.20334 | 0.183 |
| 24 | 1.19992 | 1.19972 | 1.19983 | 1.19983 | 0.39999 | 0.10666 | 0.10666 | 0.10666 | 0.050 |
| 25 | 1.19982 | 16.20072 | 1.19981 | 1.19982 | 0.39999 | 0.10666 | 0.23999 | 0.10666 | 0.056 |
| 26 | 1.19982 | 1.19981 | 1.19980 | 16.20075 | 0.39999 | 0.10666 | 0.10666 | 0.10666 | 0.087 |
| 27 | 2.38737 | 1.19895 | 0.79964 | 0.59981 | 0.39992 | 0.15993 | 0.10662 | 0.06854 | 0.030 |
| 28 | 0.79965 | 1.19900 | 2.38703 | 16.01449 | 0.39992 | 0.06854 | 0.10662 | 0.15993 | 0.101 |

| 29 | 8.19695 | 8.19684 | 8.19331 | 8.20839 | 0.57618 | 0.22121 | 0.22122 | 0.22119 | 1.340 |
| 30 | 1.20000 | 1.20000 | 1.20000 | 1.20000 | 0.40000 | 0.10667 | 0.10667 | 0.10667 | 0.183 |
| 31 | 1.20000 | 36.20000 | 1.20000 | 1.20000 | 0.40000 | 0.10667 | 0.24000 | 0.10667 | 0.298 |
| 32 | 1.20000 | 1.20000 | 1.20000 | 36.20000 | 0.40000 | 0.10667 | 0.10667 | 0.10667 | 0.549 |
| 33 | 2.40000 | 1.20000 | 0.80000 | 0.60000 | 0.40000 | 0.16000 | 0.10667 | 0.06857 | 0.146 |
| 34 | 0.80000 | 1.20000 | 2.39998 | 36.00002 | 0.40000 | 0.06857 | 0.10667 | 0.16000 | 0.682 |

Table 3 shows the percent difference between the decomposition and simulation estimates, which are displayed in Table S1 in the online supplement (Section S2), for space considerations.

**Table 3.** Percent difference in performance measure estimates of the EB policy obtained by decomposition and simulation for the 5-machine line Example 1.

| # | $\bar{y}_1$ | $\bar{y}_2$ | $\bar{y}_3$ | $\bar{y}_4$ | $\nu$ | $\theta_1$ | $\theta_2$ | $\theta_3$ |
|---|---|---|---|---|---|---|---|---|
| 1 | -0.106 | -0.148 | -0.569 | -0.710 | -0.648 | -0.483 | 2.509 | 2.703 |
| 2 | -0.776 | -1.112 | -0.618 | -0.281 | -0.087 | -1.582 | -2.120 | -0.907 |
| 3 | -1.085 | -1.442 | -0.421 | -0.125 | -0.022 | -1.802 | -2.261 | -0.624 |
| 4 | -1.649 | -1.260 | -0.175 | 0.127 | 0.003 | -2.255 | -2.055 | -0.360 |
| 5 | -0.041 | 0.121 | -0.340 | -0.551 | -0.527 | -1.061 | 3.800 | 3.836 |
| 6 | -0.156 | 0.126 | -0.172 | -0.015 | 0.003 | -1.327 | 4.143 | -3.363 |
| 7 | -0.165 | 0.096 | -0.176 | 0.000 | 0.003 | -11.188 | 13.130 | -4.474 |
| 8 | 0.045 | -0.441 | -0.346 | -0.457 | -0.386 | 0.488 | -0.035 | 1.786 |
| 9 | -0.067 | 0.020 | -0.011 | 0.064 | 0.008 | -0.559 | -0.006 | 1.075 |
| 10 | -0.006 | 0.025 | -0.041 | 0.128 | 0.010 | -0.485 | -0.006 | 0.406 |
| 11 | 0.146 | -0.084 | -0.620 | -0.739 | -0.449 | 1.028 | 5.059 | 6.379 |
| 12 | -0.117 | -0.153 | 0.058 | 0.036 | 0.040 | -0.847 | -1.066 | 0.193 |
| 13 | -0.028 | -0.014 | 0.066 | 0.022 | 0.039 | -0.826 | -0.728 | 0.180 |
| 14 | -0.332 | 0.066 | -0.073 | 0.052 | 0.003 | -1.104 | 1.535 | 4.599 |
| 15 | 0.024 | -0.152 | 0.055 | 0.041 | 0.050 | 0.354 | -0.901 | 0.176 |
| 16 | -1.003 | -1.617 | -0.400 | -0.054 | -0.007 | -1.403 | -2.757 | -1.065 |
| 17 | -0.102 | -0.030 | 0.007 | 0.075 | 0.023 | -0.047 | -0.053 | -0.007 |
| 18 | -0.010 | -0.014 | 0.054 | 0.050 | 0.054 | 0.018 | -0.046 | 0.046 |
| 19 | -0.049 | -0.071 | 0.010 | 0.078 | 0.056 | 0.015 | -0.041 | 0.031 |
| 20 | 0.018 | 0.049 | 0.032 | -0.120 | 0.089 | 0.035 | 0.010 | 0.059 |
| 21 | 0.025 | -0.021 | 0.007 | 0.052 | 0.020 | 0.013 | 0.005 | -0.033 |
| 22 | 0.047 | 0.060 | 0.051 | -0.108 | 0.091 | -0.024 | 0.039 | 0.099 |
| 23 | -0.254 | -0.104 | -0.009 | 0.187 | 0.026 | -0.063 | -0.004 | -0.026 |
| 24 | -0.159 | 0.088 | -0.176 | 0.006 | 0.003 | -0.153 | -0.104 | -0.061 |
| 25 | -0.227 | 0.023 | -0.048 | 0.128 | 0.010 | -0.048 | -0.075 | -0.034 |
| 26 | -0.140 | -0.278 | -0.184 | 0.040 | 0.039 | -0.095 | -0.016 | -0.080 |
| 27 | -0.336 | 0.070 | -0.072 | 0.053 | 0.003 | -0.014 | 0.020 | -0.112 |
| 28 | -0.024 | -0.371 | -0.401 | 0.089 | 0.038 | -0.155 | -0.079 | -0.020 |
| 29 | -0.500 | -0.269 | -0.118 | 0.514 | 0.027 | -0.086 | -0.033 | -0.033 |
| 30 | -0.165 | 0.096 | -0.179 | 0.003 | 0.003 | -0.153 | -0.104 | -0.061 |
| 31 | -0.303 | 0.027 | -0.044 | 0.130 | 0.010 | -0.071 | -0.085 | -0.034 |
| 32 | -0.219 | -0.345 | -0.244 | 0.040 | 0.039 | -0.119 | -0.039 | -0.103 |
| 33 | -0.330 | 0.076 | -0.069 | 0.053 | 0.003 | -0.012 | 0.020 | -0.109 |
| 34 | -0.158 | -0.535 | -0.567 | 0.066 | 0.040 | -0.176 | -0.101 | -0.039 |

From the results in Table 3, we make the following observations regarding the accuracy of the decomposition method with respect to simulation:

(1) In all cases, the accuracy of the decomposition method is very high. More specifically, the absolute percent difference in the average throughput estimate and average stage WIP levels does not exceed 0.7% and 1.7%, respectively. The absolute percent difference in the average overflow probabilities does not exceed 2.8%, except in cases 5, 6, 7, 11, and 14, where the overflow probabilities are negligible anyway (they are less than 0.018).

(2) The accuracy of the decomposition method in estimating the average throughput appears mostly to be increasing in the echelon buffer capacities (e.g., compare cases 1-4, 5-7, 8-10, 11-13). Most likely, this happens because when the echelon buffer capacities increase, the buffer-full and buffer-empty probabilities decrease. As a result, the decoupling effect of the buffers increases, improving the accuracy of the method. Still, the accuracy remains very high even for very low buffer capacities (cases 1, 5, 8, 11, 17-22).

(3) The accuracy of the decomposition method appears mostly to be increasing in the production probability (rate) of the machines (compare cases 3 and 16). Most likely, this happens because when the production probability of a machine increases, the variability of its processing time decreases (recall from Assumption (3) in Section 4, that the squared coefficient of variation of the processing time of machine $M_n$ is $1 - p_n$), again resulting in increasing the decoupling effect of the buffers.

(4) The accuracy of the decomposition method appears to be higher for the lines with a slower machine and the unbalanced lines than for the balanced lines (compare cases 1-3 (balanced) vs. cases 5-13 (balanced except for a slower machine), and cases 14-15 (unbalanced)). Having a slow machine in the line effectively separates the line into two segments, one upstream and the other downstream of that machine. The slow machine is almost never starved and hence almost always feeds the downstream segment independently of what is going on in the upstream segment. This decoupling effect again seems to help increase the accuracy of the decomposition method.

(5) The accuracy of the decomposition method in estimating the average throughput is higher for the unbalanced lines where the machines have increasing production probabilities moving down the line than for the lines that have decreasing probabilities (compare cases 14, 21, 27, and 33 vs. cases 15, 22, 28, and 34). In the lines where the machines have increasing production probabilities, inventory tends to decrease downstream the line and the machines are hardly ever blocked. Hence, they almost always feed their downstream segment independently of what is going on in the downstream segment. This decoupling effect again seems to help increase the accuracy of the decomposition method.

By comparing the last column of Table 2 and Table S1, we make the following observations regarding the computational efficiency of the decomposition method compared to that of simulation:

(1) The computational time using simulation is more or less the same for all cases examined. This is expected because in all cases we used time-driven simulation with the same simulation horizon.

(2) The computational time using decomposition is less than 1% of the corresponding time using simulation in most cases. In a few cases (4 and 2), it goes up to 6%.

(3) The computational time using decomposition is increasing in the echelon buffer capacities. This is expected because the larger the capacities, the larger the number of states of the Markov chain of the two-machine subsystems $\tilde{L}_n$ that need to be solved.

(4) The computational time using decomposition is smaller for the lines with a slower machine and the unbalanced lines than it is for the balanced lines (compare cases 1-2 (balanced) vs. cases 4-5, 6-7, and 8-9 (balanced except for a slower machine), and cases 10-11 (unbalanced)). Most likely, this happens because of the stronger decoupling effect in the first cases discussed earlier.

Moreover, by comparing the performance measures between the different cases in Table 2 (and Table S1), we make the following observations regarding the effect of system parameters on system performance:

(1) The average echelon WIP levels and line throughput are increasing in the intermediate buffer capacities. As the echelon buffer capacities increase, the average line throughput approaches the production probability of the slowest machine.

(2) The overflow rate is decreasing in the echelon buffer capacities.

(3) The average echelon WIP levels and line throughput is increasing in the production probability (rate) of the machines while the overflow rate is decreasing.

(4) Having a slower machine in the line results in increasing the average echelon WIP levels upstream of that machine and decreasing them downstream of the machine.

(5) A production line in which the total buffer space is allocated to the last buffer only (CONWIP) yields a higher average throughput and a lower average total WIP than the same line in which the total buffer space is evenly allocated among all intermediate buffers (e.g., compare case 3 vs. case 29, case 12 vs. case 26, etc.). The concentration of the average total WIP towards the downstream stages, however, seems to be higher in the former line than it is in the latter line, as a result of the fact that under CONWIP no machine except the first one is ever blocked. If the inventory holding cost rate is also an increasing function of the stages – a reasonable assumption, given the extra value added in each stage – then the total weighted inventory cost could end up been higher in the former line than it is in the latter line, even though the total average WIP is smaller under CONWIP. Moreover, the former line also yields higher overflow probabilities than the latter line, resulting in a higher transfer rate – and hence cost – of parts to remote buffers.

Finally, to explore the difference in performance between the EB and IB policies, we simulated the 5-machine production line under the IB policy for the 34 cases in Table 1. The performance measure estimates of that policy are shown in Table S2 in the online supplement (Section S2). In all cases, except case 11, both the average throughput and the average total stage WIP level under the EB policy are higher than their respective values under the IB policy. As expected, the biggest differences in average throughput (60% and above) are observed in cases 17-34, where the EB policy is equivalent to CONWIP. Not surprisingly, these are the cases with the biggest differences in the average total stage WIP. Case 11 is the only case where the average throughput and the average total stage WIP level under the EB policy are lower than their respective values under the IB policy. As was mentioned in Section 1, this can happen in short lines with low buffer capacities where the WIP-limit of the EB policy is significantly smaller that the WIP-limit of the IB policy. Case 11 fits this description because the WIP-limit of the EB policy is 5 whereas the WIP-limit of the IB policy is 8. However, the low buffer capacities is not the only reason that the average throughput and the average total stage WIP level under the EB policy are lower than their respective values under the IB policy in case 11. For example, note that cases 1, 5, and 8 have the exact same buffer allocation as case 11, but result in higher average throughput and average total stage WIP level under the EB policy than they do under the IB policy. The difference between these cases and case 11 is that in case 11 there is a slower machine at the end of the line. That machine seems to block more frequently the release of new parts into the line under EB, resulting in a reduced average throughput and average total stage WIP level compared to IB. Finally, recall that a disadvantage of the EB policy compared to the IB policy is that in the former policy, parts are transferred for storage to remote downstream buffers at rates equal to the overflow probabilities. This transfer may incur a cost. Under the IB policy, on the other hand, no part is ever transferred to a remote buffer.

In Example 2, we consider a production line consisting of $N = 10$ machines and 9 buffers. For this system, we evaluated 27 different instances. The rationale behind the choice of parameter values for the different instances is similar to that in Example 1. For space considerations, the input data and the results for each instance are presented in Section S3 in the online supplement. The observations on the results of Example 1 presented above still hold for the results of Example 2. One important difference is that the computational time of the decomposition method in Example 2 is higher than it is in Example 1. This is natural because in Example 2, there are twice as many stages (machines) and – more importantly – the echelon buffer capacities are much higher. Nonetheless, in most cases, the computational time of the decomposition method still remains significantly lower than the corresponding time of simulation.

## 9 Conclusions

We introduced the EB policy for controlling the flow of parts through a production line, and we developed a decomposition-based approximation method for evaluating its performance. Our numerical results indicate that this method is computationally efficient and highly accurate when compared to simulation. They also indicate that an EB policy where the entire buffer space is allocated to the last intermediate buffer (CONWIP) yields higher average throughput and lower average WIP than the same policy in which the buffer space is evenly allocated among all intermediate buffers. The tradeoff is that the concentration of the average total WIP towards the downstream stages and the overflow probabilities are higher in the former case than they are in the latter case. At the same time, the EB policy generally yields higher average throughput, at the cost of higher average WIP and overflow probabilities, than the IB policy. Based on these results, a promising direction for future research is to use the developed approximation method to optimally design the echelon buffer capacities, and compare the performance of the resulting optimal EB policy against that of the optimal IB and CONWIP policies. Another possible direction is to generalize the decomposition method for more complicated machine behavior models than the Bernoulli model. Even under the Bernoulli machine assumption, however, it would also be useful to come up with a more efficient way to analyze the two-machine subsystems in isolation in the decomposition method.

## Acknowledgements

This research was supported by the ECSEL Joint Undertaking under grant agreement No 737459. This Joint Undertaking receives support from the European Union's Horizon 2020 research and innovation program and Germany, Austria, France, Czech Republic, Netherlands, Belgium, Spain, Greece, Sweden, Italy, Ireland, Poland, Hungary, Portugal, Denmark, Finland, Luxembourg, Norway, Turkey.

Online supplement for "Performance evaluation of a production line operated under an echelon buffer policy" by George Liberopoulos

## S1  Derivation of expressions (10)-(13)

The derivation of expression (10) is as follows:

$$\lambda_n(x_n) = P(M_n \text{ produces a part}|x_n) = \frac{P(M_n \text{ produces a part}, x_n)}{P(x_n)} = \frac{\sum_{y_{n-1}=0}^{K_{n-1}-x_n} P(M_n \text{ produces a part}, y_{n-1}, x_n)}{\sum_{y_{n-1}=0}^{K_{n-1}-x_n} P_n(y_{n-1}, x_n)}$$

$$= \frac{\sum_{y_{n-1}=0}^{K_{n-1}-x_n} P(M_n \text{ produces a part}|y_{n-1}, x_n) P_n(y_{n-1}, x_n)}{\sum_{y_{n-1}=0}^{K_{n-1}-x_n} P_n(y_{n-1}, x_n)} = \frac{\sum_{y_{n-1}=0}^{K_{n-1}-x_n} P_n(y_{n-1}, x_n) p_n(y_{n-1})}{\sum_{y_{n-1}=0}^{K_{n-1}-x_n} P_n(y_{n-1}, x_n)}.$$

The derivation of expression (11) is as follows:

$$v_n(x_{n-1}) = P(\widetilde{M}_{n+1} \text{ produces a part}|x_{n-1}) = \frac{P(\widetilde{M}_{n+1} \text{ produces a part}, x_{n-1})}{P(x_{n-1})}$$

$$= \frac{P(\widetilde{M}_{n+1} \text{ produces a part}, y_{n-1} + x_n = x_{n-1})}{P(y_{n-1} + x_n = x_{n-1})}$$

$$= \frac{\sum_{y_{n-1}=(x_{n-1}-K_n)^+}^{x_{n-1}} P(\widetilde{M}_{n+1} \text{ produces a part}, y_{n-1}, x_n = x_{n-1} - y_{n-1})}{\sum_{y_{n-1}=(x_{n-1}-K_n)^+}^{x_{n-1}} P(y_{n-1}, x_n = x_{n-1} - y_{n-1})}$$

$$= \frac{\sum_{y_{n-1}=(x_{n-1}-K_n)^+}^{x_{n-1}} P(\widetilde{M}_{n+1} \text{ produces a part}|\, y_{n-1}, x_n = x_{n-1} - y_{n-1}) P(y_{n-1}, x_n = x_{n-1} - y_{n-1})}{\sum_{y_{n-1}=(x_{n-1}-K_n)^+}^{x_{n-1}} P(y_{n-1}, x_n = x_{n-1} - y_{n-1})}$$

$$= \frac{\sum_{y_{n-1}=(x_{n-1}-K_n)^+}^{x_{n-1}} P_n(y_{n-1}, x_{n-1} - y_{n-1}) q_{n+1}(x_{n-1} - y_{n-1})}{\sum_{y_{n-1}=(x_{n-1}-K_n)^+}^{x_{n-1}} P_n(y_{n-1}, x_{n-1} - y_{n-1})},$$

where we used the facts that $y_{n-1} \leq x_{n-1} \leq K_{n-1}$ and $y_{n-1} = x_{n-1} - x_n \geq x_{n-1} - K_n \geq 0$.

The derivation of expression (12) is as follows:

$$\bar{x}_n = \sum_{x_n=0}^{K_n} x_n P(x_n) = \sum_{x_n=0}^{K_n} x_n \sum_{y_{n-1}=0}^{K_{n-1}-x_n} P_n(y_{n-1}, x_n).$$

Finally, the derivation of expression (13) is as follows:

$$\theta_{n-1} = P(y_{n-1} \geq C_{n-1} + 1, Q_{n-1} \text{ receives a part}, M_{n-1} \text{ does not produce a part})$$

$$= \sum_{x_n=0}^{K_n} P(y_{n-1} \geq K_{n-1} - K_n + 1, x_n, Q_{n-1} \text{ receives a part}, M_{n-1} \text{ does not produce a part})$$

$$= \sum_{x_n=0}^{K_n} \sum_{y_{n-1}=K_{n-1}-K_n+1}^{K_{n-1}-x_n} P(y_{n-1}, x_n, Q_{n-1} \text{ receives a part}, M_{n-1} \text{ does not produce a part})$$

$$= \sum_{x_n=0}^{K_n} \sum_{y_{n-1}=K_{n-1}-K_n+1}^{K_{n-1}-x_n} P(Q_{n-1} \text{ receives a part}, M_{n-1} \text{ does not produce a part}|y_{n-1}, x_n) P_n(y_{n-1}, x_n)$$

$$= \sum_{x_n=0}^{K_n} \sum_{y_{n-1}=K_{n-1}-K_n+1}^{K_{n-1}-x_n} P_n(y_{n-1}, x_n) r_{n-1}(y_{n-1} + x_n) \bar{p}_n(y_{n-1}).$$

## S2 Performance measure estimates of the EB and IB policy for Example 1 obtained by simulation

Table S1 shows the performance measure estimates of the EB policy for Example 1 obtained by simulation. To get these estimates, for each instance, we executed 30 independent time-driven simulation runs over a horizon of 500,000 periods. For each estimate that we computed, we report the sample mean and a 95% confidence interval over the 30 runs.

**Table S1.** Performance measure estimates of the EB policy for the 5-machine line Example 1 obtained by simulation.

| # | $\bar{y}_1$ | $\bar{y}_2$ | $\bar{y}_3$ | $\bar{y}_4$ | $\nu$ | $\theta_1$ | $\theta_2$ | $\theta_3$ | CPU (s) |
|---|---|---|---|---|---|---|---|---|---|
| 1 | 1.23294 ± 0.00081 | 1.10032 ± 0.00074 | 0.97988 ± 0.00068 | 0.78038 ± 0.00051 | 0.38284 ± 0.00013 | 0.03443 ± 0.0001 | 0.02662 ± 0.00007 | 0.01831 ± 0.00006 | 46.37 |
| 2 | 5.52841 ± 0.01409 | 4.94656 ± 0.01163 | 4.28351 ± 0.00893 | 2.66693 ± 0.00397 | 0.5424 ± 0.00015 | 0.09018 ± 0.00045 | 0.07699 ± 0.0004 | 0.06066 ± 0.00036 | 47.39 |
| 3 | 10.94555 ± 0.04989 | 9.82158 ± 0.04734 | 8.46077 ± 0.03205 | 4.95836 ± 0.01547 | 0.57007 ± 0.00012 | 0.10567 ± 0.0008 | 0.09138 ± 0.00084 | 0.07339 ± 0.00061 | 47.51 |
| 4 | 16.40232 ± 0.10252 | 14.67664 ± 0.10404 | 12.63462 ± 0.07582 | 7.23345 ± 0.03451 | 0.57972 ± 0.00014 | 0.11223 ± 0.00104 | 0.09691 ± 0.00129 | 0.07854 ± 0.00105 | 47.64 |
| 5 | 0.8912 ± 0.00097 | 0.85756 ± 0.00086 | 0.80348 ± 0.00072 | 0.67005 ± 0.00051 | 0.33574 ± 0.00013 | 0.01962 ± 0.00008 | 0.01651 ± 0.00007 | 0.01214 ± 0.00005 | 46.08 |
| 6 | 1.20251 ± 0.00335 | 1.20268 ± 0.00321 | 1.22551 ± 0.00335 | 1.17611 ± 0.00294 | 0.39998 ± 0.00023 | 0.00187 ± 0.00005 | 0.00179 ± 0.00005 | 0.002 ± 0.00005 | 46.17 |
| 7 | 1.20199 ± 0.00335 | 1.19885 ± 0.00332 | 1.20261 ± 0.00318 | 1.19953 ± 0.00365 | 0.39999 ± 0.00023 | 0.00004 ± 0.00001 | 0.00003 ± 0.00001 | 0.00003 ± 0.00001 | 46.19 |
| 8 | 1.14855 ± 0.00074 | 1.73997 ± 0.00129 | 0.75753 ± 0.00072 | 0.64295 ± 0.00059 | 0.32428 ± 0.00014 | 0.0225 ± 0.00009 | 0.07937 ± 0.0001 | 0.01058 ± 0.00005 | 45.71 |
| 9 | 5.01535 ± 0.00323 | 12.39876 ± 0.00584 | 1.2165 ± 0.00266 | 1.17239 ± 0.00289 | 0.39971 ± 0.00024 | 0.0364 ± 0.00018 | 0.2372 ± 0.00015 | 0.00177 ± 0.00005 | 46.26 |
| 10 | 10.00077 ± 0.00347 | 27.39295 ± 0.0061 | 1.20097 ± 0.00296 | 1.19799 ± 0.00352 | 0.39996 ± 0.00025 | 0.0364 ± 0.00018 | 0.24001 ± 0.00013 | 0.00003 ± 0.00001 | 46.31 |
| 11 | 1.14106 ± 0.00084 | 1.05178 ± 0.00072 | 0.983 ± 0.0006 | 1.13053 ± 0.00072 | 0.31931 ± 0.00016 | 0.02147 ± 0.0001 | 0.01666 ± 0.00007 | 0.01196 ± 0.00006 | 45.63 |
| 12 | 5.01594 ± 0.0033 | 5.0011 ± 0.00287 | 4.97156 ± 0.0025 | 4.83386 ± 0.00288 | 0.39821 ± 0.00028 | 0.03609 ± 0.00019 | 0.03568 ± 0.00017 | 0.03469 ± 0.00018 | 45.84 |
| 13 | 10.00293 ± 0.00363 | 10.00122 ± 0.00308 | 9.99242 ± 0.00301 | 9.79914 ± 0.00327 | 0.39982 ± 0.00028 | 0.03651 ± 0.00019 | 0.03647 ± 0.00017 | 0.03613 ± 0.0002 | 46.03 |
| 14 | 2.39542 ± 0.01167 | 1.19809 ± 0.00329 | 0.80029 ± 0.0011 | 0.59943 ± 0.00074 | 0.39991 ± 0.00023 | 0.02112 ± 0.00032 | 0.00181 ± 0.00007 | 0.00012 ± 0.00001 | 46.26 |
| 15 | 5.18375 ± 0.00105 | 5.35391 ± 0.00251 | 5.72825 ± 0.0042 | 4.15532 ± 0.00521 | 0.39054 ± 0.00025 | 0.02326 ± 0.0001 | 0.04564 ± 0.00018 | 0.08386 ± 0.00033 | 45.74 |
| 16 | 10.9008 ± 0.05163 | 9.81577 ± 0.06259 | 8.43901 ± 0.04514 | 4.98538 ± 0.01542 | 0.77905 ± 0.00015 | 0.07285 ± 0.00063 | 0.06355 ± 0.00076 | 0.05093 ± 0.00058 | 48.60 |
| 17 | 1.00101 ± 0.00078 | 1.0003 ± 0.00083 | 0.99994 ± 0.00094 | 0.99924 ± 0.00111 | 0.4004 ± 0.00011 | 0.10079 ± 0.00013 | 0.1008 ± 0.00012 | 0.10075 ± 0.00014 | 46.52 |
| 18 | 0.78472 ± 0.00087 | 0.78476 ± 0.00082 | 0.78422 ± 0.00091 | 0.78425 ± 0.00088 | 0.34061 ± 0.00013 | 0.07322 ± 0.00013 | 0.07327 ± 0.00013 | 0.0732 ± 0.00015 | 45.99 |
| 19 | 0.78503 ± 0.00077 | 1.86277 ± 0.00176 | 0.78457 ± 0.00082 | 0.78403 ± 0.0011 | 0.3406 ± 0.00015 | 0.07322 ± 0.00011 | 0.16484 ± 0.00013 | 0.07321 ± 0.00013 | 46.00 |
| 20 | 0.7845 ± 0.00085 | 0.78427 ± 0.00091 | 0.78437 ± 0.0009 | 1.86367 ± 0.00198 | 0.34049 ± 0.00017 | 0.07321 ± 0.0001 | 0.07323 ± 0.00013 | 0.07319 ± 0.00013 | 45.57 |
| 21 | 1.16001 ± 0.00153 | 0.80406 ± 0.00083 | 0.60782 ± 0.00058 | 0.48594 ± 0.00033 | 0.34518 ± 0.00011 | 0.1134 ± 0.00015 | 0.0756 ± 0.00013 | 0.04862 ± 0.00011 | 46.18 |

| # | | | | | | | | | |
|---|---|---|---|---|---|---|---|---|---|
| 22 | 0.60759 ± 0.00058 | 0.80339 ± 0.00099 | 1.15975 ± 0.00191 | 1.94382 ± 0.0023 | 0.34494 ± 0.00017 | 0.04862 ± 0.00011 | 0.07558 ± 0.00015 | 0.1133 ± 0.00021 | 45.73 |
| 23 | 4.20999 ± 0.01481 | 4.20558 ± 0.01684 | 4.19844 ± 0.01761 | 4.19302 ± 0.01681 | 0.55267 ± 0.00015 | 0.20348 ± 0.00021 | 0.20337 ± 0.00022 | 0.20339 ± 0.00031 | 46.47 |
| 24 | 1.20182 ± 0.00336 | 1.19866 ± 0.0033 | 1.20194 ± 0.00318 | 1.19975 ± 0.0037 | 0.39998 ± 0.00023 | 0.10683 ± 0.00022 | 0.10677 ± 0.00022 | 0.10673 ± 0.00021 | 46.01 |
| 25 | 1.20255 ± 0.00239 | 16.19698 ± 0.00756 | 1.20039 ± 0.00293 | 1.19829 ± 0.00356 | 0.39995 ± 0.00025 | 0.10671 ± 0.00014 | 0.24017 ± 0.00013 | 0.1067 ± 0.00019 | 45.64 |
| 26 | 1.2015 ± 0.00279 | 1.20315 ± 0.00311 | 1.202 ± 0.00374 | 16.19432 ± 0.01025 | 0.39984 ± 0.00028 | 0.10676 ± 0.00017 | 0.10668 ± 0.00024 | 0.10675 ± 0.00024 | 45.56 |
| 27 | 2.39539 ± 0.01168 | 1.19811 ± 0.00329 | 0.80022 ± 0.00109 | 0.5995 ± 0.00075 | 0.39991 ± 0.00023 | 0.15995 ± 0.00027 | 0.1066 ± 0.0002 | 0.06862 ± 0.00015 | 45.82 |
| 28 | 0.79985 ± 0.00145 | 1.20345 ± 0.00341 | 2.39661 ± 0.01664 | 16.0003 ± 0.01953 | 0.39977 ± 0.00028 | 0.06865 ± 0.00016 | 0.1067 ± 0.00025 | 0.15996 ± 0.0004 | 45.54 |
| 29 | 8.23794 ± 0.05831 | 8.21891 ± 0.07001 | 8.203 ± 0.06019 | 8.16617 ± 0.06291 | 0.57603 ± 0.00013 | 0.2214 ± 0.00021 | 0.2213 ± 0.00023 | 0.22127 ± 0.00031 | 47.46 |
| 30 | 1.20199 ± 0.00335 | 1.19884 ± 0.00332 | 1.20215 ± 0.00317 | 1.19996 ± 0.00369 | 0.39999 ± 0.00023 | 0.10683 ± 0.00022 | 0.10678 ± 0.00022 | 0.10673 ± 0.00021 | 46.33 |
| 31 | 1.20363 ± 0.00242 | 36.19007 ± 0.00756 | 1.20053 ± 0.00296 | 1.19844 ± 0.00354 | 0.39996 ± 0.00025 | 0.10674 ± 0.00014 | 0.2402 ± 0.00013 | 0.1067 ± 0.00019 | 46.36 |
| 32 | 1.20263 ± 0.00283 | 1.20413 ± 0.00309 | 1.20293 ± 0.00387 | 36.18559 ± 0.01051 | 0.39984 ± 0.00028 | 0.10679 ± 0.00017 | 0.10671 ± 0.00024 | 0.10678 ± 0.00024 | 46.04 |
| 33 | 2.40792 ± 0.01239 | 1.19908 ± 0.00333 | 0.80055 ± 0.00108 | 0.59968 ± 0.00075 | 0.39999 ± 0.00023 | 0.16002 ± 0.00027 | 0.10664 ± 0.0002 | 0.06865 ± 0.00015 | 46.46 |
| 34 | 0.80126 ± 0.00154 | 1.20642 ± 0.0035 | 2.4136 ± 0.01795 | 35.97615 ± 0.02084 | 0.39984 ± 0.00028 | 0.06869 ± 0.00016 | 0.10677 ± 0.00025 | 0.16006 ± 0.0004 | 46.10 |

From the results in Table S1, we observe that in all cases, the confidence intervals of the performance measures obtained by simulation are quite tight. More specifically, the confidence intervals of the throughput estimates are below 0.15% of these estimates. The confidence intervals of the average echelon WIP level estimates are looser but still remain below 1% of these estimates in all cases except for a few cases (4, 16, 28, 29, 33, and 34) where they are below 1.8%. Finally, the confidence intervals for the overflow probabilities remain well below 3% of these estimates for most cases, except cases 6, 7, 9, 10, and 14 where the estimates themselves are extremely low.

**Table S2.** Performance measure estimates of the IB policy for the 5-machine line Example 1 obtained by simulation.

| # | $\bar{y}_1$ | $\bar{y}_2$ | $\bar{y}_3$ | $\bar{y}_4$ | $\nu$ | CPU (s) |
|---|---|---|---|---|---|---|
| 1 | 1.2335 ± 0.0006 | 1.0665 ± 0.0005 | 0.9329 ± 0.0006 | 0.7665 ± 0.0005 | 0.3774 ± 0.0001 | 32.741 |
| 2 | 3.6273 ± 0.0042 | 3.1777 ± 0.0053 | 2.8175 ± 0.0049 | 2.3716 ± 0.0047 | 0.5257 ± 0.0001 | 32.889 |
| 3 | 6.6471 ± 0.0117 | 5.8218 ± 0.0186 | 5.1576 ± 0.0148 | 4.3507 ± 0.0162 | 0.56 ± 0.0001 | 32.958 |
| 4 | 9.67 ± 0.0267 | 8.4583 ± 0.037 | 7.4886 ± 0.0351 | 6.3167 ± 0.0358 | 0.5726 ± 0.0001 | 32.927 |
| 5 | 0.837 ± 0.0008 | 0.7988 ± 0.0008 | 0.7451 ± 0.0007 | 0.6462 ± 0.0005 | 0.3259 ± 0.0001 | 32.528 |
| 6 | 1.1745 ± 0.0026 | 1.1821 ± 0.0026 | 1.1816 ± 0.0027 | 1.1603 ± 0.0028 | 0.3983 ± 0.0002 | 32.697 |
| 7 | 1.2009 ± 0.0034 | 1.1984 ± 0.0032 | 1.2016 ± 0.0033 | 1.1992 ± 0.0037 | 0.4 ± 0.0002 | 32.749 |
| 8 | 1.3847 ± 0.0006 | 1.3065 ± 0.0005 | 0.6927 ± 0.0006 | 0.6156 ± 0.0006 | 0.3131 ± 0.0001 | 32.284 |
| 9 | 4.8514 ± 0.0023 | 4.8405 ± 0.0022 | 1.156 ± 0.0021 | 1.1456 ± 0.0028 | 0.3969 ± 0.0002 | 32.603 |
| 10 | 9.797 ± 0.0029 | 9.7986 ± 0.0027 | 1.1991 ± 0.0029 | 1.1972 ± 0.0036 | 0.3999 ± 0.0002 | 32.513 |
| 11 | 1.3542 ± 0.0007 | 1.2549 ± 0.0007 | 1.2015 ± 0.0009 | 1.1637 ± 0.0008 | 0.3258 ± 0.0001 | 32.421 |
| 12 | 4.8397 ± 0.0029 | 4.8184 ± 0.0031 | 4.816 ± 0.0032 | 4.8245 ± 0.003 | 0.3981 ± 0.0003 | 32.583 |
| 13 | 9.7989 ± 0.0036 | 9.7975 ± 0.0032 | 9.7951 ± 0.004 | 9.7989 ± 0.0034 | 0.3998 ± 0.0003 | 32.604 |
| 14 | 1.8544 ± 0.0039 | 1.0966 ± 0.0021 | 0.7617 ± 0.001 | 0.5787 ± 0.0006 | 0.3905 ± 0.0002 | 32.664 |
| 15 | 5.4213 ± 0.0007 | 5.2389 ± 0.0012 | 4.9017 ± 0.0025 | 4.1441 ± 0.0054 | 0.3904 ± 0.0002 | 32.550 |
| 16 | 6.6036 ± 0.0177 | 5.8173 ± 0.0206 | 5.1738 ± 0.0248 | 4.4018 ± 0.0139 | 0.7716 ± 0.0001 | 33.025 |
| 17 | 0.596 ± 0.0003 | 0.5 ± 0.0003 | 0.4039 ± 0.0002 | 0.4354 ± 0.0004 | 0.2425 ± 0.0001 | 32.275 |
| 18 | 0.4698 ± 0.0003 | 0.4203 ± 0.0004 | 0.3532 ± 0.0002 | 0.3746 ± 0.0004 | 0.2121 ± 0.0001 | 31.951 |
| 19 | 0.6582 ± 0.0003 | 0.599 ± 0.0003 | 0.3419 ± 0.0003 | 0.3605 ± 0.0004 | 0.2052 ± 0.0001 | 31.723 |
| 20 | 0.5962 ± 0.0003 | 0.5004 ± 0.0003 | 0.4046 ± 0.0002 | 0.8526 ± 0.0015 | 0.2423 ± 0.0001 | 32.161 |
| 21 | 0.4945 ± 0.0003 | 0.3687 ± 0.0004 | 0.2888 ± 0.0002 | 0.2549 ± 0.0002 | 0.2023 ± 0.0001 | 31.718 |
| 22 | 0.6869 ± 0.0002 | 0.6154 ± 0.0003 | 0.5024 ± 0.0003 | 0.914 ± 0.0017 | 0.2505 ± 0.0001 | 32.080 |
| 23 | 0.596 ± 0.0003 | 0.5 ± 0.0003 | 0.4039 ± 0.0002 | 0.4354 ± 0.0004 | 0.2425 ± 0.0001 | 32.699 |
| 24 | 0.4698 ± 0.0003 | 0.4203 ± 0.0004 | 0.3532 ± 0.0002 | 0.3746 ± 0.0004 | 0.2121 ± 0.0001 | 32.624 |
| 25 | 0.6582 ± 0.0003 | 0.599 ± 0.0003 | 0.3419 ± 0.0003 | 0.3605 ± 0.0004 | 0.2052 ± 0.0001 | 32.356 |
| 26 | 0.596 ± 0.0003 | 0.5 ± 0.0003 | 0.4039 ± 0.0002 | 0.8585 ± 0.0017 | 0.2425 ± 0.0001 | 32.761 |
| 27 | 0.4945 ± 0.0003 | 0.3687 ± 0.0004 | 0.2888 ± 0.0002 | 0.2549 ± 0.0002 | 0.2023 ± 0.0001 | 32.466 |
| 28 | 0.6865 ± 0.0002 | 0.6149 ± 0.0003 | 0.5015 ± 0.0003 | 0.9259 ± 0.0019 | 0.2508 ± 0.0001 | 32.663 |
| 29 | 0.596 ± 0.0003 | 0.5 ± 0.0003 | 0.4039 ± 0.0002 | 0.4354 ± 0.0004 | 0.2425 ± 0.0001 | 32.484 |
| 30 | 0.4698 ± 0.0003 | 0.4203 ± 0.0004 | 0.3532 ± 0.0002 | 0.3746 ± 0.0004 | 0.2121 ± 0.0001 | 32.078 |
| 31 | 0.6582 ± 0.0003 | 0.599 ± 0.0003 | 0.3419 ± 0.0003 | 0.3605 ± 0.0004 | 0.2052 ± 0.0001 | 31.703 |
| 32 | 0.596 ± 0.0003 | 0.5 ± 0.0003 | 0.4039 ± 0.0002 | 0.8585 ± 0.0017 | 0.2425 ± 0.0001 | 32.244 |
| 33 | 0.4945 ± 0.0003 | 0.3687 ± 0.0004 | 0.2888 ± 0.0002 | 0.2549 ± 0.0002 | 0.2023 ± 0.0001 | 31.841 |
| 34 | 0.6865 ± 0.0002 | 0.6149 ± 0.0003 | 0.5015 ± 0.0003 | 0.9259 ± 0.0019 | 0.2508 ± 0.0001 | 32.094 |

## S3 Input data and performance measure estimates of the EB and IB policies for Example 2

Table S3 shows the input data for each instance of Example 2.

Table S3. Input data for the 10-machine line Example 2.

| # | $p_1$ | $p_2$ | $p_3$ | $p_4$ | $p_5$ | $p_6$ | $p_7$ | $p_8$ | $p_9$ | $p_{10}$ | $C_1$ | $C_2$ | $C_3$ | $C_4$ | $C_5$ | $C_6$ | $C_7$ | $C_8$ | $C_9$ | $K_1$ | $K_2$ | $K_3$ | $K_4$ | $K_5$ | $K_6$ | $K_7$ | $K_8$ | $K_9$ |
|---|---|---|---|---|---|---|---|---|---|---|---|---|---|---|---|---|---|---|---|---|---|---|---|---|---|---|---|---|
| 1 | 0.6 | 0.6 | 0.6 | 0.6 | 0.6 | 0.6 | 0.6 | 0.6 | 0.6 | 0.6 | 1 | 1 | 1 | 1 | 1 | 1 | 1 | 1 | 1 | 10 | 9 | 8 | 7 | 6 | 5 | 4 | 3 | 2 |
| 2 | 0.6 | 0.6 | 0.6 | 0.6 | 0.6 | 0.6 | 0.6 | 0.6 | 0.6 | 0.6 | 5 | 5 | 5 | 5 | 5 | 5 | 5 | 5 | 5 | 46 | 41 | 36 | 31 | 26 | 21 | 16 | 11 | 6 |
| 3 | 0.6 | 0.6 | 0.6 | 0.6 | 0.6 | 0.6 | 0.6 | 0.6 | 0.6 | 0.6 | 10 | 10 | 10 | 10 | 10 | 10 | 10 | 10 | 10 | 91 | 81 | 71 | 61 | 51 | 41 | 31 | 21 | 11 |
| 4 | 0.6 | **0.4** | 0.6 | 0.6 | 0.6 | 0.6 | 0.6 | 0.6 | 0.6 | 0.6 | 1 | 1 | 1 | 1 | 1 | 1 | 1 | 1 | 1 | 10 | 9 | 8 | 7 | 6 | 5 | 4 | 3 | 2 |
| 5 | 0.6 | **0.4** | 0.6 | 0.6 | 0.6 | 0.6 | 0.6 | 0.6 | 0.6 | 0.6 | 5 | 5 | 5 | 5 | 5 | 5 | 5 | 5 | 5 | 46 | 41 | 36 | 31 | 26 | 21 | 16 | 11 | 6 |
| 6 | 0.6 | **0.4** | 0.6 | 0.6 | 0.6 | 0.6 | 0.6 | 0.6 | 0.6 | 0.6 | 10 | 10 | 10 | 10 | 10 | 10 | 10 | 10 | 10 | 91 | 81 | 71 | 61 | 51 | 41 | 31 | 21 | 11 |
| 7 | 0.6 | 0.6 | 0.6 | 0.6 | 0.6 | **0.4** | 0.6 | 0.6 | 0.6 | 0.6 | 1 | 1 | 1 | 1 | 1 | 1 | 1 | 1 | 1 | 10 | 9 | 8 | 7 | 6 | 5 | 4 | 3 | 2 |
| 8 | 0.6 | 0.6 | 0.6 | 0.6 | 0.6 | **0.4** | 0.6 | 0.6 | 0.6 | 0.6 | 5 | 5 | 5 | 5 | 5 | 5 | 5 | 5 | 5 | 46 | 41 | 36 | 31 | 26 | 21 | 16 | 11 | 6 |
| 9 | 0.6 | 0.6 | 0.6 | 0.6 | 0.6 | **0.4** | 0.6 | 0.6 | 0.6 | 0.6 | 10 | 10 | 10 | 10 | 10 | 10 | 10 | 10 | 10 | 91 | 81 | 71 | 61 | 51 | 41 | 31 | 21 | 11 |
| 10 | 0.6 | 0.6 | 0.6 | 0.6 | 0.6 | 0.6 | 0.6 | 0.6 | **0.4** | 0.6 | 1 | 1 | 1 | 1 | 1 | 1 | 1 | 1 | 1 | 10 | 9 | 8 | 7 | 6 | 5 | 4 | 3 | 2 |
| 11 | 0.6 | 0.6 | 0.6 | 0.6 | 0.6 | 0.6 | 0.6 | 0.6 | **0.4** | 0.6 | 5 | 5 | 5 | 5 | 5 | 5 | 5 | 5 | 5 | 46 | 41 | 36 | 31 | 26 | 21 | 16 | 11 | 6 |
| 12 | 0.6 | 0.6 | 0.6 | 0.6 | 0.6 | 0.6 | 0.6 | 0.6 | **0.4** | 0.6 | 10 | 10 | 10 | 10 | 10 | 10 | 10 | 10 | 10 | 91 | 81 | 71 | 61 | 51 | 41 | 31 | 21 | 11 |
| 13 | 0.4 | 0.45 | 0.5 | 0.55 | 0.6 | 0.65 | 0.7 | 0.75 | 0.8 | 0.85 | 5 | 5 | 5 | 5 | 5 | 5 | 5 | 5 | 5 | 46 | 41 | 36 | 31 | 26 | 21 | 16 | 11 | 6 |
| 14 | 0.85 | 0.8 | 0.75 | 0.7 | 0.65 | 0.6 | 0.55 | 0.5 | 0.45 | 0.4 | 5 | 5 | 5 | 5 | 5 | 5 | 5 | 5 | 5 | 46 | 41 | 36 | 31 | 26 | 21 | 16 | 11 | 6 |
| 15 | 0.8 | 0.8 | 0.8 | 0.8 | 0.8 | 0.8 | 0.8 | 0.8 | 0.8 | 0.8 | 10 | 10 | 10 | 10 | 10 | 10 | 10 | 10 | 10 | 91 | 81 | 71 | 61 | 51 | 41 | 31 | 21 | 11 |
| 16 | 0.6 | 0.6 | 0.6 | 0.6 | 0.6 | 0.6 | 0.6 | 0.6 | 0.6 | 0.6 | 0 | 0 | 0 | 0 | 0 | 0 | 0 | 0 | 9 | 10 | 10 | 10 | 10 | 10 | 10 | 10 | 10 | 10 |
| 17 | 0.6 | **0.4** | 0.6 | 0.6 | 0.6 | 0.6 | 0.6 | 0.6 | 0.6 | 0.6 | 0 | 0 | 0 | 0 | 0 | 0 | 0 | 0 | 9 | 10 | 10 | 10 | 10 | 10 | 10 | 10 | 10 | 10 |
| 18 | 0.6 | 0.6 | 0.6 | 0.6 | 0.6 | **0.4** | 0.6 | 0.6 | 0.6 | 0.6 | 0 | 0 | 0 | 0 | 0 | 0 | 0 | 0 | 9 | 10 | 10 | 10 | 10 | 10 | 10 | 10 | 10 | 10 |
| 19 | 0.6 | 0.6 | 0.6 | 0.6 | 0.6 | 0.6 | 0.6 | 0.6 | **0.4** | 0.6 | 0 | 0 | 0 | 0 | 0 | 0 | 0 | 0 | 9 | 10 | 10 | 10 | 10 | 10 | 10 | 10 | 10 | 10 |
| 20 | 0.4 | 0.45 | 0.5 | 0.55 | 0.6 | 0.65 | 0.7 | 0.75 | 0.8 | 0.85 | 0 | 0 | 0 | 0 | 0 | 0 | 0 | 0 | 9 | 10 | 10 | 10 | 10 | 10 | 10 | 10 | 10 | 10 |
| 21 | 0.85 | 0.8 | 0.75 | 0.7 | 0.65 | 0.6 | 0.55 | 0.5 | 0.45 | 0.4 | 0 | 0 | 0 | 0 | 0 | 0 | 0 | 0 | 9 | 10 | 10 | 10 | 10 | 10 | 10 | 10 | 10 | 10 |
| 22 | 0.6 | 0.6 | 0.6 | 0.6 | 0.6 | 0.6 | 0.6 | 0.6 | 0.6 | 0.6 | 0 | 0 | 0 | 0 | 0 | 0 | 0 | 0 | 45 | 46 | 46 | 46 | 46 | 46 | 46 | 46 | 46 | 46 |
| 23 | 0.6 | **0.4** | 0.6 | 0.6 | 0.6 | 0.6 | 0.6 | 0.6 | 0.6 | 0.6 | 0 | 0 | 0 | 0 | 0 | 0 | 0 | 0 | 45 | 46 | 46 | 46 | 46 | 46 | 46 | 46 | 46 | 46 |
| 24 | 0.6 | 0.6 | 0.6 | 0.6 | 0.6 | **0.4** | 0.6 | 0.6 | 0.6 | 0.6 | 0 | 0 | 0 | 0 | 0 | 0 | 0 | 0 | 45 | 46 | 46 | 46 | 46 | 46 | 46 | 46 | 46 | 46 |
| 25 | 0.6 | 0.6 | 0.6 | 0.6 | 0.6 | 0.6 | 0.6 | 0.6 | **0.4** | 0.6 | 0 | 0 | 0 | 0 | 0 | 0 | 0 | 0 | 45 | 46 | 46 | 46 | 46 | 46 | 46 | 46 | 46 | 46 |
| 26 | 0.4 | 0.45 | 0.5 | 0.55 | 0.6 | 0.65 | 0.7 | 0.75 | 0.8 | 0.85 | 0 | 0 | 0 | 0 | 0 | 0 | 0 | 0 | 45 | 46 | 46 | 46 | 46 | 46 | 46 | 46 | 46 | 46 |
| 27 | 0.85 | 0.8 | 0.75 | 0.7 | 0.65 | 0.6 | 0.55 | 0.5 | 0.45 | 0.4 | 0 | 0 | 0 | 0 | 0 | 0 | 0 | 0 | 45 | 46 | 46 | 46 | 46 | 46 | 46 | 46 | 46 | 46 |

Tables S4-S6 show the performance measure estimates of the EB policy obtained by decomposition and simulation, and the percent difference between these estimates, for Example 2. Finally, Table S7 shows the performance measure estimates of the IB policy obtained by simulation for the same example. As in the case of Example 1, for the simulation results, we report the sample mean and a 95% confidence interval based on 30 independent time-driven simulation runs over a horizon of 500,000 periods.

Table S4. Performance measure estimates of the EB policy for the 10-machine line Example 2 obtained by decomposition.

| # | $\bar{y}_1$ | $\bar{y}_2$ | $\bar{y}_3$ | $\bar{y}_4$ | $\bar{y}_5$ | $\bar{y}_6$ | $\bar{y}_7$ | $\bar{y}_8$ | $\bar{y}_9$ | $v$ |
|---|---|---|---|---|---|---|---|---|---|---|
| 1 | 1.2531 | 1.1642 | 1.1104 | 1.0683 | 1.0293 | 0.9885 | 0.9384 | 0.8636 | 0.7080 | 0.3522 |
| 2 | 5.7921 | 5.4224 | 5.2129 | 5.0493 | 4.8920 | 4.7121 | 4.4658 | 4.0234 | 2.5889 | 0.5376 |
| 3 | 11.4298 | 10.7553 | 10.3609 | 10.0515 | 9.7541 | 9.4113 | 8.9265 | 8.0223 | 4.8469 | 0.5681 |
| 4 | 2.4936 | 0.9019 | 0.9030 | 0.8966 | 0.8847 | 0.8660 | 0.8364 | 0.7828 | 0.6544 | 0.3287 |
| 5 | 35.1950 | 1.2001 | 1.1999 | 1.2000 | 1.2002 | 1.2008 | 1.2044 | 1.2236 | 1.1760 | 0.4000 |
| 6 | 80.2000 | 1.2000 | 1.2000 | 1.2000 | 1.2000 | 1.2000 | 1.2000 | 1.2005 | 1.1995 | 0.4000 |
| 7 | 1.1874 | 1.1120 | 1.0691 | 1.0373 | 1.8884 | 0.7990 | 0.7856 | 0.7447 | 0.6300 | 0.3182 |
| 8 | 5.0137 | 5.0012 | 5.0001 | 5.0000 | 19.9810 | 1.2006 | 1.2042 | 1.2234 | 1.1759 | 0.4000 |
| 9 | 10.0002 | 10.0000 | 10.0000 | 10.0000 | 44.9998 | 1.2000 | 1.2000 | 1.2005 | 1.1995 | 0.4000 |

| # | | | | | | | | | | |
|---|---|---|---|---|---|---|---|---|---|---|
| 10 | 1.1700 | 1.0991 | 1.0598 | 1.0313 | 1.0069 | 0.9825 | 0.9542 | 1.3844 | 0.6030 | 0.3074 |
| 11 | 5.0130 | 5.0010 | 4.9999 | 4.9995 | 4.9988 | 4.9971 | 4.9924 | 8.6573 | 1.1525 | 0.3984 |
| 12 | 10.0002 | 10.0000 | 10.0000 | 10.0000 | 10.0000 | 9.9999 | 9.9998 | 18.6018 | 1.1985 | 0.4000 |
| 13 | 4.7779 | 2.3976 | 1.5993 | 1.1996 | 0.9598 | 0.7998 | 0.6856 | 0.5999 | 0.5333 | 0.4000 |
| 14 | 5.0594 | 5.0753 | 5.0986 | 5.1344 | 5.1929 | 5.2940 | 5.4668 | 5.5838 | 3.5948 | 0.3792 |
| 15 | 11.3999 | 10.7339 | 10.3465 | 10.0423 | 9.7478 | 9.4069 | 8.9271 | 8.0189 | 4.8845 | 0.7777 |
| 16 | 1.0000 | 1.0000 | 1.0000 | 1.0000 | 1.0000 | 1.0000 | 1.0000 | 1.0000 | 1.0000 | 0.3832 |
| 17 | 2.4337 | 0.8406 | 0.8407 | 0.8407 | 0.8407 | 0.8408 | 0.8408 | 0.8407 | 0.8407 | 0.3450 |
| 18 | 0.8407 | 0.8407 | 0.8407 | 0.8407 | 2.4338 | 0.8405 | 0.8408 | 0.8407 | 0.8407 | 0.3450 |
| 19 | 0.8407 | 0.8407 | 0.8407 | 0.8407 | 0.8407 | 0.8407 | 0.8407 | 2.4337 | 0.8407 | 0.3450 |
| 20 | 1.6804 | 1.2658 | 1.0070 | 0.8327 | 0.7083 | 0.6154 | 0.5436 | 0.4866 | 0.4403 | 0.3422 |
| 21 | 0.4866 | 0.5437 | 0.6154 | 0.7083 | 0.8327 | 1.0070 | 1.2657 | 1.6805 | 2.4197 | 0.3422 |
| 22 | 4.6012 | 4.6000 | 4.6001 | 4.5994 | 4.5989 | 4.5982 | 4.6011 | 4.6007 | 4.5988 | 0.5516 |
| 23 | 35.2000 | 1.2000 | 1.2000 | 1.2000 | 1.2000 | 1.2000 | 1.2000 | 1.2000 | 1.2000 | 0.4000 |
| 24 | 1.2000 | 1.2000 | 1.2000 | 1.2000 | 35.2000 | 1.2000 | 1.2000 | 1.2000 | 1.2000 | 0.4000 |
| 25 | 1.2000 | 1.2000 | 1.2000 | 1.2000 | 1.2000 | 1.2000 | 1.2000 | 35.2000 | 1.2000 | 0.4000 |
| 26 | 4.7778 | 2.3977 | 1.5993 | 1.1996 | 0.9598 | 0.7998 | 0.6856 | 0.5999 | 0.5333 | 0.4000 |
| 27 | 0.5999 | 0.6856 | 0.7999 | 0.9598 | 1.1997 | 1.5994 | 2.3982 | 4.7750 | 32.4492 | 0.4000 |

| # | $\theta_1$ | $\theta_2$ | $\theta_3$ | $\theta_4$ | $\theta_5$ | $\theta_6$ | $\theta_7$ | $\theta_8$ | CPU (s) |
|---|---|---|---|---|---|---|---|---|---|
| 1 | 0.0340 | 0.0307 | 0.0288 | 0.0271 | 0.0254 | 0.0232 | 0.0202 | 0.0146 | 0.080 |
| 2 | 0.0914 | 0.0832 | 0.0789 | 0.0757 | 0.0727 | 0.0694 | 0.0644 | 0.0538 | 1.449 |
| 3 | 0.1071 | 0.0977 | 0.0930 | 0.0896 | 0.0865 | 0.0830 | 0.0780 | 0.0665 | 13.113 |
| 4 | 0.1198 | 0.0215 | 0.0212 | 0.0206 | 0.0198 | 0.0185 | 0.0163 | 0.0121 | 0.018 |
| 5 | 0.2400 | 0.0018 | 0.0018 | 0.0018 | 0.0019 | 0.0019 | 0.0019 | 0.0019 | 0.387 |
| 6 | 0.2400 | 0.0000 | 0.0000 | 0.0000 | 0.0000 | 0.0000 | 0.0000 | 0.0000 | 1.785 |
| 7 | 0.0260 | 0.0235 | 0.0220 | 0.0208 | 0.0893 | 0.0158 | 0.0143 | 0.0108 | 0.026 |
| 8 | 0.0363 | 0.0360 | 0.0360 | 0.0360 | 0.2399 | 0.0018 | 0.0019 | 0.0019 | 0.475 |
| 9 | 0.0362 | 0.0362 | 0.0362 | 0.0362 | 0.2400 | 0.0000 | 0.0000 | 0.0000 | 2.339 |
| 10 | 0.0239 | 0.0215 | 0.0202 | 0.0191 | 0.0180 | 0.0166 | 0.0146 | 0.0477 | 0.030 |
| 11 | 0.0359 | 0.0356 | 0.0356 | 0.0356 | 0.0355 | 0.0355 | 0.0353 | 0.2205 | 0.480 |
| 12 | 0.0362 | 0.0362 | 0.0362 | 0.0362 | 0.0362 | 0.0362 | 0.0362 | 0.2394 | 2.348 |
| 13 | 0.0701 | 0.0210 | 0.0063 | 0.0018 | 0.0005 | 0.0001 | 0.0000 | 0.0000 | 0.365 |
| 14 | 0.0085 | 0.0133 | 0.0193 | 0.0272 | 0.0373 | 0.0504 | 0.0679 | 0.0882 | 0.200 |
| 15 | 0.0743 | 0.0678 | 0.0645 | 0.0621 | 0.0600 | 0.0576 | 0.0541 | 0.0461 | 13.848 |
| 16 | 0.0949 | 0.0949 | 0.0949 | 0.0949 | 0.0949 | 0.0949 | 0.0949 | 0.0949 | 0.055 |
| 17 | 0.1737 | 0.0772 | 0.0772 | 0.0772 | 0.0772 | 0.0772 | 0.0772 | 0.0772 | 0.049 |
| 18 | 0.0772 | 0.0772 | 0.0772 | 0.0772 | 0.1737 | 0.0772 | 0.0772 | 0.0772 | 0.047 |
| 19 | 0.0772 | 0.0772 | 0.0772 | 0.0772 | 0.0772 | 0.0772 | 0.0772 | 0.1737 | 0.063 |
| 20 | 0.1396 | 0.1142 | 0.0934 | 0.0761 | 0.0615 | 0.0489 | 0.0381 | 0.0286 | 0.035 |
| 21 | 0.0286 | 0.0381 | 0.0489 | 0.0615 | 0.0761 | 0.0934 | 0.1142 | 0.1396 | 0.051 |
| 22 | 0.2026 | 0.2026 | 0.2026 | 0.2026 | 0.2026 | 0.2026 | 0.2026 | 0.2026 | 8.882 |
| 23 | 0.2400 | 0.1067 | 0.1067 | 0.1067 | 0.1067 | 0.1067 | 0.1067 | 0.1067 | 1.415 |
| 24 | 0.1067 | 0.1067 | 0.1067 | 0.1067 | 0.2400 | 0.1067 | 0.1067 | 0.1067 | 2.243 |
| 25 | 0.1067 | 0.1067 | 0.1067 | 0.1067 | 0.1067 | 0.1067 | 0.1067 | 0.2400 | 4.023 |
| 26 | 0.1955 | 0.1600 | 0.1309 | 0.1066 | 0.0861 | 0.0686 | 0.0533 | 0.0400 | 1.051 |
| 27 | 0.0400 | 0.0533 | 0.0686 | 0.0861 | 0.1066 | 0.1309 | 0.1600 | 0.1955 | 5.030 |

**Table S5.** Performance measure estimates of the EB policy for the 10-machine line Example 2 obtained by simulation.

| # | $\bar{y}_1$ | $\bar{y}_2$ | $\bar{y}_3$ | $\bar{y}_4$ | $\bar{y}_5$ | $\bar{y}_6$ | $\bar{y}_7$ | $\bar{y}_8$ | $\bar{y}_9$ | $\nu$ |
|---|---|---|---|---|---|---|---|---|---|---|
| 1 | 1.2304 | 1.1433 | 1.0987 | 1.0673 | 1.0383 | 1.0056 | 0.9611 | 0.8873 | 0.7258 | 0.3599 |
| | ± 0.0012 | ± 0.0009 | ± 0.0012 | ± 0.0012 | ± 0.001 | ± 0.0009 | ± 0.0009 | ± 0.0007 | ± 0.0004 | ± 0.0001 |
| 2 | 5.6988 | 5.3539 | 5.1908 | 5.0836 | 4.9716 | 4.8332 | 4.5953 | 4.1098 | 2.6157 | 0.5392 |
| | ± 0.0184 | ± 0.0149 | ± 0.0158 | ± 0.0187 | ± 0.0155 | ± 0.015 | ± 0.011 | ± 0.0101 | ± 0.004 | ± 0.0001 |
| 3 | 11.285 | 10.6162 | 10.3114 | 10.0861 | 9.9063 | 9.6779 | 9.1972 | 8.1635 | 4.8897 | 0.5688 |
| | ± 0.0639 | ± 0.0602 | ± 0.0551 | ± 0.0723 | ± 0.0547 | ± 0.0576 | ± 0.0414 | ± 0.0368 | ± 0.0112 | ± 0.0001 |
| 4 | 2.4851 | 0.8959 | 0.895 | 0.892 | 0.8859 | 0.8734 | 0.8496 | 0.7985 | 0.6671 | 0.3344 |
| | ± 0.0029 | ± 0.001 | ± 0.001 | ± 0.0009 | ± 0.0009 | ± 0.0009 | ± 0.0008 | ± 0.0008 | ± 0.0005 | ± 0.0001 |
| 5 | 35.1924 | 1.1996 | 1.1978 | 1.2004 | 1.2006 | 1.201 | 1.204 | 1.2248 | 1.1758 | 0.3999 |
| | ± 0.0156 | ± 0.0032 | ± 0.0026 | ± 0.0034 | ± 0.003 | ± 0.0027 | ± 0.0022 | ± 0.0033 | ± 0.0019 | ± 0.0002 |
| 6 | 80.1707 | 1.1996 | 1.1978 | 1.2004 | 1.2005 | 1.2002 | 1.1996 | 1.2015 | 1.1993 | 0.3999 |
| | ± 0.0153 | ± 0.0032 | ± 0.0026 | ± 0.0034 | ± 0.003 | ± 0.0027 | ± 0.0022 | ± 0.0032 | ± 0.0023 | ± 0.0002 |
| 7 | 1.1671 | 1.0929 | 1.0578 | 1.0355 | 1.9196 | 0.8101 | 0.7973 | 0.7587 | 0.6417 | 0.3234 |
| | ± 0.0012 | ± 0.0008 | ± 0.001 | ± 0.0009 | ± 0.0017 | ± 0.0006 | ± 0.0008 | ± 0.0006 | ± 0.0003 | ± 0.0001 |
| 8 | 5.0156 | 5.0001 | 4.9967 | 4.9993 | 19.974 | 1.2016 | 1.2043 | 1.2259 | 1.1771 | 0.4 |
| | ± 0.0039 | ± 0.0027 | ± 0.0034 | ± 0.0035 | ± 0.0086 | ± 0.0025 | ± 0.0029 | ± 0.0027 | ± 0.0023 | ± 0.0002 |
| 9 | 10.0006 | 9.996 | 9.9928 | 9.9955 | 44.9778 | 1.201 | 1.2003 | 1.2031 | 1.2007 | 0.4001 |
| | ± 0.0043 | ± 0.0031 | ± 0.0038 | ± 0.0039 | ± 0.0088 | ± 0.0024 | ± 0.0028 | ± 0.0024 | ± 0.0029 | ± 0.0002 |
| 10 | 1.146 | 1.0771 | 1.0457 | 1.0288 | 1.0132 | 0.9988 | 0.9773 | 1.4199 | 0.6143 | 0.3125 |
| | ± 0.0009 | ± 0.0008 | ± 0.0009 | ± 0.0009 | ± 0.001 | ± 0.0008 | ± 0.0007 | ± 0.0012 | ± 0.0005 | ± 0.0001 |
| 11 | 5.0133 | 4.9992 | 4.9968 | 4.9997 | 4.9998 | 5.0003 | 4.9969 | 8.6571 | 1.1516 | 0.3982 |
| | ± 0.0036 | ± 0.003 | ± 0.0037 | ± 0.0036 | ± 0.0031 | ± 0.0031 | ± 0.0025 | ± 0.0042 | ± 0.0023 | ± 0.0003 |
| 12 | 9.9998 | 9.9955 | 9.9927 | 9.9965 | 9.9958 | 9.9965 | 9.9959 | 18.5967 | 1.1975 | 0.3998 |
| | ± 0.0042 | ± 0.0035 | ± 0.0041 | ± 0.0038 | ± 0.0034 | ± 0.0033 | ± 0.0031 | ± 0.0049 | ± 0.0028 | ± 0.0003 |
| 13 | 4.7984 | 2.3888 | 1.5996 | 1.2004 | 0.9609 | 0.8006 | 0.6857 | 0.6004 | 0.5336 | 0.4 |
| | ± 0.048 | ± 0.0106 | ± 0.0044 | ± 0.0028 | ± 0.0016 | ± 0.0012 | ± 0.001 | ± 0.001 | ± 0.0006 | ± 0.0002 |
| 14 | 5.0578 | 5.0728 | 5.0941 | 5.1293 | 5.1841 | 5.2881 | 5.4892 | 5.5902 | 3.5933 | 0.379 |
| | ± 0.0006 | ± 0.0009 | ± 0.0011 | ± 0.0016 | ± 0.0026 | ± 0.0031 | ± 0.0041 | ± 0.0048 | ± 0.0045 | ± 0.0002 |
| 15 | 11.2683 | 10.6334 | 10.307 | 10.0811 | 9.8786 | 9.6301 | 9.2027 | 8.1835 | 4.8993 | 0.778 |
| | ± 0.0644 | ± 0.083 | ± 0.0895 | ± 0.0906 | ± 0.0638 | ± 0.0685 | ± 0.046 | ± 0.0519 | ± 0.0148 | ± 0.0001 |
| 16 | 1.0004 | 0.9999 | 0.9996 | 1.0003 | 1.0009 | 1.0002 | 1 | 0.9996 | 0.9998 | 0.3831 |
| | ± 0.0012 | ± 0.0011 | ± 0.0016 | ± 0.0015 | ± 0.0013 | ± 0.0012 | ± 0.0013 | ± 0.0014 | ± 0.0012 | ± 0.0001 |
| 17 | 2.4336 | 0.84 | 0.8407 | 0.8407 | 0.8414 | 0.8412 | 0.8409 | 0.8407 | 0.8405 | 0.345 |
| | ± 0.0032 | ± 0.0011 | ± 0.0012 | ± 0.0011 | ± 0.0009 | ± 0.0011 | ± 0.001 | ± 0.0012 | ± 0.001 | ± 0.0001 |
| 18 | 0.8415 | 0.8407 | 0.8403 | 0.8413 | 2.432 | 0.8415 | 0.8411 | 0.8409 | 0.8407 | 0.345 |
| | ± 0.0012 | ± 0.001 | ± 0.0013 | ± 0.0012 | ± 0.0033 | ± 0.0009 | ± 0.0012 | ± 0.001 | ± 0.0009 | ± 0.0001 |
| 19 | 0.8407 | 0.8405 | 0.84 | 0.8406 | 0.8413 | 0.8409 | 0.8406 | 2.4352 | 0.8398 | 0.3449 |
| | ± 0.0012 | ± 0.0012 | ± 0.0012 | ± 0.001 | ± 0.0012 | ± 0.0009 | ± 0.0009 | ± 0.0037 | ± 0.0011 | ± 0.0002 |
| 20 | 1.6802 | 1.2654 | 1.0063 | 0.8328 | 0.7089 | 0.6156 | 0.5435 | 0.4868 | 0.4403 | 0.342 |
| | ± 0.0032 | ± 0.0024 | ± 0.0014 | ± 0.0011 | ± 0.0009 | ± 0.0005 | ± 0.0005 | ± 0.0006 | ± 0.0004 | ± 0.0001 |
| 21 | 0.4867 | 0.5437 | 0.6145 | 0.7084 | 0.8325 | 1.0058 | 1.2649 | 1.6832 | 2.4201 | 0.342 |
| | ± 0.0004 | ± 0.0006 | ± 0.0006 | ± 0.0008 | ± 0.0012 | ± 0.0017 | ± 0.0013 | ± 0.0033 | ± 0.0042 | ± 0.0001 |
| 22 | 4.6254 | 4.604 | 4.5961 | 4.5961 | 4.5998 | 4.6017 | 4.5914 | 4.6039 | 4.581 | 0.5515 |
| | ± 0.0256 | ± 0.0241 | ± 0.0227 | ± 0.027 | ± 0.0221 | ± 0.0224 | ± 0.027 | ± 0.0235 | ± 0.0166 | ± 0.0001 |
| 23 | 35.1959 | 1.1996 | 1.1978 | 1.2004 | 1.2005 | 1.2002 | 1.1996 | 1.2011 | 1.1998 | 0.3999 |
| | ± 0.0156 | ± 0.0032 | ± 0.0026 | ± 0.0034 | ± 0.003 | ± 0.0027 | ± 0.0022 | ± 0.0032 | ± 0.0023 | ± 0.0002 |
| 24 | 1.2047 | 1.2023 | 1.2005 | 1.2032 | 35.1772 | 1.201 | 1.2003 | 1.2027 | 1.201 | 0.4001 |
| | ± 0.0036 | ± 0.0027 | ± 0.0033 | ± 0.0031 | ± 0.0145 | ± 0.0024 | ± 0.0028 | ± 0.0024 | ± 0.003 | ± 0.0002 |
| 25 | 1.2026 | 1.201 | 1.1986 | 1.2008 | 1.2008 | 1.2016 | 1.1998 | 35.1888 | 1.1988 | 0.3998 |
| | ± 0.0035 | ± 0.0039 | ± 0.0028 | ± 0.0027 | ± 0.0036 | ± 0.0027 | ± 0.0023 | ± 0.0173 | ± 0.0029 | ± 0.0003 |
| 26 | 4.7984 | 2.3888 | 1.5996 | 1.2004 | 0.9609 | 0.8006 | 0.6857 | 0.6004 | 0.5336 | 0.4 |
| | ± 0.048 | ± 0.0106 | ± 0.0044 | ± 0.0028 | ± 0.0016 | ± 0.0012 | ± 0.001 | ± 0.001 | ± 0.0006 | ± 0.0002 |
| 27 | 0.6009 | 0.6866 | 0.7995 | 0.9621 | 1.2021 | 1.6008 | 2.4005 | 4.8086 | 32.4026 | 0.3998 |
| | ± 0.0007 | ± 0.001 | ± 0.0013 | ± 0.0023 | ± 0.0026 | ± 0.0043 | ± 0.0103 | ± 0.0487 | ± 0.0597 | ± 0.0002 |

| # | $\theta_1$ | $\theta_2$ | $\theta_3$ | $\theta_4$ | $\theta_5$ | $\theta_6$ | $\theta_7$ | $\theta_8$ | CPU (s) |
|---|---|---|---|---|---|---|---|---|---|
| 1 | 0.0321 ± 0.0001 | 0.0281 ± 0.0001 | 0.0266 ± 0.0001 | 0.0255 ± 0.0001 | 0.0244 ± 0.0001 | 0.0228 ± 0.0001 | 0.0201 ± 0.0001 | 0.0147 ± 0.0001 | 51.958 |
| 2 | 0.0908 ± 0.0005 | 0.0826 ± 0.0004 | 0.0792 ± 0.0005 | 0.077 ± 0.0006 | 0.075 ± 0.0004 | 0.0724 ± 0.0005 | 0.0675 ± 0.0004 | 0.056 ± 0.0004 | 52.283 |
| 3 | 0.107 ± 0.0009 | 0.0973 ± 0.0008 | 0.0937 ± 0.0008 | 0.091 ± 0.0011 | 0.0891 ± 0.0008 | 0.0872 ± 0.0009 | 0.082 ± 0.0008 | 0.0686 ± 0.0007 | 52.467 |
| 4 | 0.1211 ± 0.0002 | 0.0213 ± 0.0001 | 0.0203 ± 0.0001 | 0.0196 ± 0.0001 | 0.019 ± 0.0001 | 0.0179 ± 0.0001 | 0.016 ± 0.0001 | 0.012 ± 0.0001 | 54.509 |
| 5 | 0.24 ± 0.0001 | 0.0018 ± 0.0001 | 0.0018 ± 0.0001 | 0.0019 ± 0.0001 | 0.0018 ± 0.0001 | 0.0019 ± 0 | 0.0018 ± 0.0001 | 0.002 ± 0.0001 | 51.111 |
| 6 | 0.24 ± 0.0001 | 0 ± 0 | 0 ± 0 | 0 ± 0 | 0 ± 0 | 0 ± 0 | 0 ± 0 | 0 ± 0 | 50.647 |
| 7 | 0.0242 ± 0.0001 | 0.0208 ± 0.0001 | 0.0196 ± 0.0001 | 0.0188 ± 0.0001 | 0.091 ± 0.0001 | 0.0158 ± 0.0001 | 0.014 ± 0.0001 | 0.0106 ± 0 | 50.831 |
| 8 | 0.0365 ± 0.0002 | 0.0359 ± 0.0001 | 0.0359 ± 0.0002 | 0.0359 ± 0.0002 | 0.24 ± 0.0001 | 0.0019 ± 0.0001 | 0.0019 ± 0.0001 | 0.002 ± 0 | 51.215 |
| 9 | 0.0364 ± 0.0002 | 0.0362 ± 0.0001 | 0.0362 ± 0.0002 | 0.0362 ± 0.0002 | 0.2401 ± 0.0001 | 0 ± 0 | 0 ± 0 | 0 ± 0 | 51.168 |
| 10 | 0.0214 ± 0.0001 | 0.0184 ± 0.0001 | 0.0174 ± 0.0001 | 0.0169 ± 0.0001 | 0.0161 ± 0.0001 | 0.0155 ± 0.0001 | 0.014 ± 0 | 0.0493 ± 0.0001 | 53.672 |
| 11 | 0.0358 ± 0.0002 | 0.0353 ± 0.0002 | 0.0354 ± 0.0002 | 0.0354 ± 0.0002 | 0.0355 ± 0.0002 | 0.0355 ± 0.0002 | 0.0353 ± 0.0001 | 0.2203 ± 0.0002 | 53.230 |
| 12 | 0.0363 ± 0.0002 | 0.0361 ± 0.0002 | 0.0361 ± 0.0002 | 0.0362 ± 0.0002 | 0.0362 ± 0.0002 | 0.0363 ± 0.0002 | 0.0362 ± 0.0001 | 0.2392 ± 0.0001 | 53.119 |
| 13 | 0.0705 ± 0.0007 | 0.0208 ± 0.0003 | 0.0063 ± 0.0001 | 0.0018 ± 0.0001 | 0.0005 ± 0 | 0.0001 ± 0 | 0 ± 0 | 0 ± 0 | 53.284 |
| 14 | 0.0083 ± 0.0001 | 0.0129 ± 0.0001 | 0.0189 ± 0.0001 | 0.0266 ± 0.0002 | 0.0366 ± 0.0002 | 0.05 ± 0.0002 | 0.0688 ± 0.0002 | 0.0883 ± 0.0003 | 54.235 |
| 15 | 0.074 ± 0.0006 | 0.0681 ± 0.0008 | 0.0652 ± 0.0009 | 0.0634 ± 0.0008 | 0.0618 ± 0.0006 | 0.06 ± 0.0007 | 0.0571 ± 0.0006 | 0.0481 ± 0.0007 | 55.286 |
| 16 | 0.0949 ± 0.0002 | 0.0949 ± 0.0001 | 0.0949 ± 0.0002 | 0.0949 ± 0.0002 | 0.095 ± 0.0002 | 0.0949 ± 0.0001 | 0.0949 ± 0.0002 | 0.0949 ± 0.0001 | 51.195 |
| 17 | 0.1737 ± 0.0002 | 0.0771 ± 0.0002 | 0.0772 ± 0.0001 | 0.0772 ± 0.0002 | 0.0773 ± 0.0001 | 0.0772 ± 0.0001 | 0.0773 ± 0.0001 | 0.0772 ± 0.0001 | 50.651 |
| 18 | 0.0773 ± 0.0002 | 0.0772 ± 0.0001 | 0.0772 ± 0.0001 | 0.0773 ± 0.0002 | 0.1737 ± 0.0002 | 0.0773 ± 0.0001 | 0.0773 ± 0.0002 | 0.0772 ± 0.0001 | 50.706 |
| 19 | 0.0772 ± 0.0002 | 0.0772 ± 0.0001 | 0.0772 ± 0.0001 | 0.0772 ± 0.0002 | 0.0773 ± 0.0001 | 0.0772 ± 0.0001 | 0.0773 ± 0.0001 | 0.1738 ± 0.0002 | 51.151 |
| 20 | 0.1395 ± 0.0002 | 0.1141 ± 0.0002 | 0.0934 ± 0.0001 | 0.0761 ± 0.0002 | 0.0615 ± 0.0001 | 0.049 ± 0.0001 | 0.0381 ± 0.0001 | 0.0286 ± 0.0001 | 50.918 |
| 21 | 0.0286 ± 0.0001 | 0.038 ± 0.0001 | 0.0489 ± 0.0001 | 0.0615 ± 0.0001 | 0.0761 ± 0.0002 | 0.0933 ± 0.0002 | 0.1141 ± 0.0001 | 0.1396 ± 0.0002 | 50.197 |
| 22 | 0.2029 ± 0.0002 | 0.2027 ± 0.0002 | 0.2026 ± 0.0003 | 0.2027 ± 0.0003 | 0.2026 ± 0.0002 | 0.2027 ± 0.0002 | 0.2027 ± 0.0003 | 0.2027 ± 0.0002 | 52.986 |
| 23 | 0.24 ± 0.0001 | 0.1066 ± 0.0002 | 0.1065 ± 0.0002 | 0.1066 ± 0.0002 | 0.1067 ± 0.0002 | 0.1066 ± 0.0002 | 0.1067 ± 0.0002 | 0.1067 ± 0.0002 | 51.629 |
| 24 | 0.1069 ± 0.0002 | 0.1068 ± 0.0002 | 0.1068 ± 0.0002 | 0.1068 ± 0.0002 | 0.2401 ± 0.0001 | 0.1068 ± 0.0001 | 0.1067 ± 0.0002 | 0.1068 ± 0.0002 | 51.466 |
| 25 | 0.1067 ± 0.0002 | 0.1066 ± 0.0002 | 0.1066 ± 0.0002 | 0.1066 ± 0.0002 | 0.1068 ± 0.0002 | 0.1068 ± 0.0002 | 0.1067 ± 0.0002 | 0.24 ± 0.0002 | 51.687 |
| 26 | 0.1956 ± 0.0003 | 0.1599 ± 0.0003 | 0.1309 ± 0.0002 | 0.1067 ± 0.0002 | 0.0862 ± 0.0002 | 0.0686 ± 0.0001 | 0.0534 ± 0.0001 | 0.0401 ± 0.0002 | 51.930 |

| 27 | 0.0401 | 0.0533 | 0.0686 | 0.0862 | 0.1066 | 0.1309 | 0.16 | 0.1955 | 55.114 |
|---|---|---|---|---|---|---|---|---|---|
|  | ± 0.0001 | ± 0.0001 | ± 0.0002 | ± 0.0002 | ± 0.0002 | ± 0.0002 | ± 0.0003 | ± 0.0003 |  |

**Table S6.** Percent difference in performance measure estimates of the EB policy obtained by decomposition and simulation for the 10-machine line Example 2.

| # | $\bar{y}_1$ | $\bar{y}_2$ | $\bar{y}_3$ | $\bar{y}_4$ | $\bar{y}_5$ | $\bar{y}_6$ | $\bar{y}_7$ | $\bar{y}_8$ | $\bar{y}_9$ | $\nu$ |
|---|---|---|---|---|---|---|---|---|---|---|
| 1 | 1.808 | 1.797 | 1.061 | 0.096 | -0.871 | -1.728 | -2.413 | -2.742 | -2.522 | -2.189 |
| 2 | 1.611 | 1.262 | 0.424 | -0.679 | -1.626 | -2.569 | -2.900 | -2.147 | -1.034 | -0.295 |
| 3 | 1.267 | 1.293 | 0.478 | -0.344 | -1.561 | -2.832 | -3.033 | -1.760 | -0.882 | -0.117 |
| 4 | 0.340 | 0.662 | 0.881 | 0.515 | -0.143 | -0.856 | -1.575 | -2.006 | -1.941 | -1.720 |
| 5 | 0.007 | 0.043 | 0.174 | -0.034 | -0.034 | -0.011 | 0.030 | -0.096 | 0.021 | 0.016 |
| 6 | 0.036 | 0.034 | 0.181 | -0.035 | -0.039 | -0.017 | 0.036 | -0.089 | 0.017 | 0.016 |
| 7 | 1.707 | 1.723 | 1.062 | 0.168 | -1.649 | -1.391 | -1.492 | -1.876 | -1.854 | -1.632 |
| 8 | -0.039 | 0.021 | 0.069 | 0.013 | 0.035 | -0.086 | -0.004 | -0.206 | -0.096 | -0.014 |
| 9 | -0.005 | 0.040 | 0.072 | 0.045 | 0.049 | -0.087 | -0.026 | -0.214 | -0.094 | -0.014 |
| 10 | 2.050 | 2.003 | 1.325 | 0.241 | -0.633 | -1.663 | -2.431 | -2.563 | -1.876 | -1.644 |
| 11 | -0.007 | 0.035 | 0.061 | -0.005 | -0.020 | -0.064 | -0.089 | 0.003 | 0.079 | 0.038 |
| 12 | 0.004 | 0.045 | 0.073 | 0.035 | 0.042 | 0.034 | 0.039 | 0.027 | 0.087 | 0.040 |
| 13 | -0.428 | 0.366 | -0.019 | -0.060 | -0.114 | -0.089 | -0.019 | -0.080 | -0.064 | 0.002 |
| 14 | 0.030 | 0.049 | 0.088 | 0.099 | 0.170 | 0.111 | -0.409 | -0.115 | 0.043 | 0.057 |
| 15 | 1.155 | 0.936 | 0.382 | -0.386 | -1.343 | -2.373 | -3.087 | -2.052 | -0.304 | -0.048 |
| 16 | -0.044 | 0.011 | 0.041 | -0.037 | -0.090 | -0.018 | -0.003 | 0.036 | 0.018 | 0.025 |
| 17 | 0.005 | 0.061 | 0.004 | 0.000 | -0.090 | -0.057 | -0.017 | 0.002 | 0.024 | 0.015 |
| 18 | -0.089 | 0.000 | 0.043 | -0.069 | 0.074 | -0.114 | -0.037 | -0.025 | -0.001 | 0.005 |
| 19 | 0.008 | 0.031 | 0.083 | 0.011 | -0.074 | -0.029 | 0.008 | -0.063 | 0.115 | 0.033 |
| 20 | 0.011 | 0.032 | 0.072 | -0.007 | -0.077 | -0.021 | 0.021 | -0.027 | -0.008 | 0.032 |
| 21 | -0.004 | -0.013 | 0.143 | -0.015 | 0.033 | 0.118 | 0.062 | -0.158 | -0.015 | 0.041 |
| 22 | -0.526 | -0.087 | 0.087 | 0.072 | -0.019 | -0.074 | 0.211 | -0.070 | 0.389 | 0.012 |
| 23 | 0.012 | 0.035 | 0.180 | -0.035 | -0.039 | -0.017 | 0.036 | -0.088 | 0.015 | 0.016 |
| 24 | -0.395 | -0.195 | -0.039 | -0.268 | 0.065 | -0.087 | -0.026 | -0.222 | -0.087 | -0.014 |
| 25 | -0.214 | -0.083 | 0.114 | -0.070 | -0.064 | -0.137 | 0.019 | 0.032 | 0.097 | 0.040 |
| 26 | -0.430 | 0.370 | -0.020 | -0.060 | -0.114 | -0.089 | -0.019 | -0.080 | -0.064 | 0.002 |
| 27 | -0.169 | -0.139 | 0.041 | -0.237 | -0.203 | -0.087 | -0.097 | -0.702 | 0.144 | 0.033 |

| # | $\theta_1$ | $\theta_2$ | $\theta_3$ | $\theta_4$ | $\theta_5$ | $\theta_6$ | $\theta_7$ | $\theta_8$ |
|---|---|---|---|---|---|---|---|---|
| 1 | 5.503 | 8.390 | 7.553 | 5.802 | 3.656 | 1.863 | 0.641 | -0.523 |
| 2 | 0.652 | 0.691 | -0.392 | -1.668 | -3.028 | -4.437 | -4.815 | -4.005 |
| 3 | 0.092 | 0.435 | -0.762 | -1.596 | -3.016 | -4.987 | -5.205 | -3.052 |
| 4 | -1.051 | 1.073 | 4.064 | 5.085 | 4.084 | 3.125 | 2.027 | 0.670 |
| 5 | 0.009 | 0.466 | 1.859 | -0.787 | 0.387 | -0.420 | 2.034 | -1.532 |
| 6 | -0.007 | 7.101 | 5.854 | -3.291 | 3.775 | -3.706 | 21.027 | -0.337 |
| 7 | 7.195 | 11.467 | 11.106 | 9.465 | -1.857 | -0.391 | 1.972 | 1.172 |
| 8 | -0.492 | 0.220 | 0.257 | 0.098 | -0.027 | -0.110 | -0.355 | -2.068 |
| 9 | -0.579 | 0.051 | 0.102 | -0.025 | -0.033 | -4.953 | 2.739 | -4.267 |
| 10 | 10.182 | 14.629 | 14.203 | 11.757 | 10.337 | 6.618 | 4.155 | -3.323 |
| 11 | 0.254 | 0.781 | 0.639 | 0.381 | 0.201 | -0.114 | -0.187 | 0.082 |
| 12 | -0.123 | 0.362 | 0.212 | -0.026 | 0.030 | -0.143 | 0.089 | 0.076 |
| 13 | -0.572 | 1.127 | -0.286 | 0.396 | -1.194 | -1.919 | 11.483 | -20.761 |
| 14 | 2.646 | 2.572 | 2.439 | 2.027 | 1.730 | 0.871 | -1.229 | -0.113 |
| 15 | 0.415 | -0.419 | -1.077 | -2.081 | -3.083 | -4.163 | -5.683 | -4.315 |
| 16 | -0.021 | -0.003 | -0.026 | -0.007 | -0.058 | 0.000 | -0.030 | 0.007 |
| 17 | -0.003 | 0.073 | 0.045 | 0.007 | -0.083 | -0.029 | -0.091 | 0.029 |
| 18 | -0.118 | -0.047 | -0.050 | -0.063 | 0.016 | -0.139 | -0.095 | -0.045 |
| 19 | -0.038 | 0.065 | 0.029 | 0.025 | -0.124 | -0.053 | -0.086 | -0.048 |
| 20 | 0.077 | 0.114 | 0.094 | 0.088 | -0.018 | -0.033 | -0.070 | -0.230 |
| 21 | -0.169 | 0.085 | 0.107 | -0.019 | 0.032 | 0.149 | 0.073 | -0.030 |
| 22 | -0.106 | -0.038 | 0.001 | -0.045 | -0.004 | -0.059 | -0.038 | -0.054 |
| 23 | 0.003 | 0.050 | 0.112 | 0.035 | 0.005 | 0.024 | -0.061 | -0.037 |
| 24 | -0.189 | -0.085 | -0.096 | -0.113 | -0.050 | -0.113 | -0.042 | -0.102 |
| 25 | -0.060 | 0.023 | 0.061 | 0.020 | -0.118 | -0.078 | -0.028 | 0.018 |
| 26 | -0.027 | 0.050 | 0.014 | -0.013 | -0.109 | -0.110 | -0.151 | -0.267 |
| 27 | -0.252 | 0.022 | -0.018 | -0.130 | 0.014 | 0.019 | -0.007 | -0.007 |

**Table S7.** Performance measure estimates of the IB policy for the 10-machine line Example 2 obtained by simulation.

| # | $\bar{y}_1$ | $\bar{y}_2$ | $\bar{y}_3$ | $\bar{y}_4$ | $\bar{y}_5$ | $\bar{y}_6$ | $\bar{y}_7$ | $\bar{y}_8$ | $\bar{y}_9$ | $\nu$ | CPU (s) |
|---|---|---|---|---|---|---|---|---|---|---|---|
| 1 | 1.3024 ± 0.0006 | 1.18 ± 0.0006 | 1.1067 ± 0.0009 | 1.0504 ± 0.001 | 0.9994 ± 0.0009 | 0.9486 ± 0.0009 | 0.8924 ± 0.001 | 0.8195 ± 0.0008 | 0.6975 ± 0.0006 | 0.3483 ± 0.0001 | 37.250 |
| 2 | 3.812 ± 0.0041 | 3.4888 ± 0.0054 | 3.2916 ± 0.0066 | 3.1372 ± 0.007 | 2.9994 ± 0.0067 | 2.8628 ± 0.0053 | 2.707 ± 0.006 | 2.5082 ± 0.0053 | 2.1878 ± 0.0046 | 0.5126 ± 0.0001 | 37.369 |
| 3 | 6.9832 ± 0.0154 | 6.3868 ± 0.0176 | 6.0247 ± 0.0229 | 5.7409 ± 0.0286 | 5.48 ± 0.0294 | 5.2357 ± 0.0237 | 4.9562 ± 0.0196 | 4.6002 ± 0.0159 | 4.0163 ± 0.0121 | 0.5529 ± 0.0001 | 37.556 |
| 4 | 1.3921 ± 0.0006 | 0.7759 ± 0.0006 | 0.78 ± 0.0009 | 0.7765 ± 0.0009 | 0.7678 ± 0.0008 | 0.7535 ± 0.0008 | 0.7313 ± 0.0009 | 0.6929 ± 0.0007 | 0.6104 ± 0.0005 | 0.3102 ± 0.0001 | 37.079 |
| 5 | 4.8599 ± 0.0023 | 1.1588 ± 0.0026 | 1.1672 ± 0.0024 | 1.1712 ± 0.0029 | 1.1728 ± 0.0025 | 1.1734 ± 0.0022 | 1.1741 ± 0.0017 | 1.1724 ± 0.0026 | 1.1518 ± 0.0018 | 0.397 ± 0.0002 | 37.451 |
| 6 | 9.8004 ± 0.003 | 1.1982 ± 0.0033 | 1.197 ± 0.0027 | 1.1997 ± 0.0033 | 1.1998 ± 0.003 | 1.1994 ± 0.0027 | 1.1991 ± 0.0022 | 1.2006 ± 0.0032 | 1.1988 ± 0.0023 | 0.3999 ± 0.0002 | 37.301 |
| 7 | 1.4011 ± 0.0006 | 1.3247 ± 0.0006 | 1.293 ± 0.0008 | 1.279 ± 0.0008 | 1.2769 ± 0.0007 | 0.7125 ± 0.0006 | 0.703 ± 0.0006 | 0.6736 ± 0.0005 | 0.5984 ± 0.0004 | 0.3053 ± 0.0001 | 37.085 |
| 8 | 4.8498 ± 0.0022 | 4.8291 ± 0.0024 | 4.8294 ± 0.0022 | 4.8313 ± 0.0024 | 4.8405 ± 0.0025 | 1.1593 ± 0.0019 | 1.167 ± 0.0023 | 1.1695 ± 0.0022 | 1.1504 ± 0.0021 | 0.397 ± 0.0002 | 37.278 |
| 9 | 9.7989 ± 0.0026 | 9.7958 ± 0.0029 | 9.798 ± 0.0027 | 9.7987 ± 0.0029 | 9.7994 ± 0.0032 | 1.1997 ± 0.0024 | 1.1993 ± 0.0028 | 1.2018 ± 0.0025 | 1.2 ± 0.0029 | 0.4 ± 0.0002 | 37.150 |
| 10 | 1.3898 ± 0.0006 | 1.3066 ± 0.0007 | 1.2682 ± 0.0009 | 1.2462 ± 0.001 | 1.2325 ± 0.001 | 1.2239 ± 0.001 | 1.2203 ± 0.0009 | 1.224 ± 0.0007 | 0.6079 ± 0.0005 | 0.3102 ± 0.0001 | 37.136 |
| 11 | 4.8484 ± 0.0021 | 4.8273 ± 0.003 | 4.8257 ± 0.003 | 4.8272 ± 0.0025 | 4.8271 ± 0.0025 | 4.8293 ± 0.0026 | 4.8304 ± 0.0026 | 4.8404 ± 0.0023 | 1.1389 ± 0.0022 | 0.3969 ± 0.0003 | 37.105 |
| 12 | 9.7988 ± 0.0028 | 9.7953 ± 0.0034 | 9.7961 ± 0.0037 | 9.7982 ± 0.0028 | 9.7962 ± 0.0029 | 9.7983 ± 0.0032 | 9.7972 ± 0.0032 | 9.7997 ± 0.0028 | 1.197 ± 0.0029 | 0.3998 ± 0.0003 | 37.378 |
| 13 | 2.4704 ± 0.0052 | 1.7026 ± 0.0036 | 1.2883 ± 0.0023 | 1.0269 ± 0.0017 | 0.8498 ± 0.0011 | 0.7224 ± 0.0009 | 0.6273 ± 0.0008 | 0.5547 ± 0.0008 | 0.4966 ± 0.0005 | 0.3776 ± 0.0002 | 37.179 |
| 14 | 5.5039 ± 0.0005 | 5.4457 ± 0.0005 | 5.3728 ± 0.001 | 5.2784 ± 0.0009 | 5.1517 ± 0.0013 | 4.9748 ± 0.0017 | 4.7116 ± 0.0026 | 4.2965 ± 0.0033 | 3.5277 ± 0.005 | 0.3774 ± 0.0002 | 37.117 |
| 15 | 6.9338 ± 0.0139 | 6.3633 ± 0.0214 | 6.0143 ± 0.0248 | 5.742 ± 0.0332 | 5.4944 ± 0.0304 | 5.2673 ± 0.0306 | 4.9999 ± 0.0231 | 4.6519 ± 0.0221 | 4.0722 ± 0.0119 | 0.7662 ± 0.0001 | 38.473 |
| 16 | 0.6512 ± 0.0003 | 0.5883 ± 0.0003 | 0.5478 ± 0.0004 | 0.5151 ± 0.0005 | 0.4847 ± 0.0004 | 0.452 ± 0.0003 | 0.4115 ± 0.0004 | 0.3488 ± 0.0002 | 0.3685 ± 0.0004 | 0.2093 ± 0.0001 | 36.728 |
| 17 | 0.6805 ± 0.0003 | 0.451 ± 0.0003 | 0.4428 ± 0.0005 | 0.4316 ± 0.0005 | 0.4173 ± 0.0004 | 0.3984 ± 0.0004 | 0.3702 ± 0.0004 | 0.3196 ± 0.0003 | 0.3353 ± 0.0004 | 0.1918 ± 0.0001 | 36.475 |
| 18 | 0.6858 ± 0.0003 | 0.6379 ± 0.0003 | 0.6126 ± 0.0004 | 0.5978 ± 0.0005 | 0.5891 ± 0.0004 | 0.3836 ± 0.0003 | 0.3609 ± 0.0003 | 0.3143 ± 0.0002 | 0.3287 ± 0.0003 | 0.1885 ± 0.0001 | 36.518 |
| 19 | 0.6717 ± 0.0003 | 0.6174 ± 0.0003 | 0.5856 ± 0.0005 | 0.5629 ± 0.0005 | 0.5446 ± 0.0005 | 0.5278 ± 0.0004 | 0.5114 ± 0.0004 | 0.4926 ± 0.0004 | 0.3451 ± 0.0004 | 0.197 ± 0.0001 | 36.477 |

| | | | | | | | | | | |
|---|---|---|---|---|---|---|---|---|---|---|
| 20 | 0.5587 | 0.447 | 0.3795 | 0.3324 | 0.2968 | 0.2687 | 0.2449 | 0.2208 | 0.2084 | 0.1765 | 36.458 |
| | ± 0.0003 | ± 0.0003 | ± 0.0004 | ± 0.0003 | ± 0.0002 | ± 0.0002 | ± 0.0002 | ± 0.0002 | ± 0.0002 | ± 0.0001 | |
| 21 | 0.7668 | 0.7438 | 0.7205 | 0.6928 | 0.6578 | 0.6121 | 0.5474 | 0.4409 | 0.6148 | 0.1983 | 36.304 |
| | ± 0.0001 | ± 0.0002 | ± 0.0002 | ± 0.0003 | ± 0.0003 | ± 0.0003 | ± 0.0004 | ± 0.0003 | ± 0.001 | ± 0.0001 | |
| 22 | 0.6512 | 0.5883 | 0.5478 | 0.5151 | 0.4847 | 0.452 | 0.4115 | 0.3488 | 0.3685 | 0.2093 | 37.359 |
| | ± 0.0003 | ± 0.0003 | ± 0.0004 | ± 0.0005 | ± 0.0004 | ± 0.0003 | ± 0.0004 | ± 0.0002 | ± 0.0004 | ± 0.0001 | |
| 23 | 0.6805 | 0.451 | 0.4428 | 0.4316 | 0.4173 | 0.3984 | 0.3702 | 0.3196 | 0.3353 | 0.1918 | 37.250 |
| | ± 0.0003 | ± 0.0003 | ± 0.0005 | ± 0.0005 | ± 0.0004 | ± 0.0004 | ± 0.0004 | ± 0.0003 | ± 0.0004 | ± 0.0001 | |
| 24 | 0.6858 | 0.6379 | 0.6126 | 0.5978 | 0.5891 | 0.3836 | 0.3609 | 0.3143 | 0.3287 | 0.1885 | 37.214 |
| | ± 0.0003 | ± 0.0003 | ± 0.0004 | ± 0.0005 | ± 0.0004 | ± 0.0003 | ± 0.0003 | ± 0.0002 | ± 0.0003 | ± 0.0001 | |
| 25 | 0.6717 | 0.6174 | 0.5856 | 0.5629 | 0.5446 | 0.5278 | 0.5114 | 0.4926 | 0.3451 | 0.197 | 37.294 |
| | ± 0.0003 | ± 0.0003 | ± 0.0005 | ± 0.0005 | ± 0.0005 | ± 0.0004 | ± 0.0004 | ± 0.0004 | ± 0.0004 | ± 0.0001 | |
| 26 | 0.5587 | 0.447 | 0.3795 | 0.3324 | 0.2968 | 0.2687 | 0.2449 | 0.2208 | 0.2084 | 0.1765 | 36.772 |
| | ± 0.0003 | ± 0.0003 | ± 0.0004 | ± 0.0003 | ± 0.0002 | ± 0.0002 | ± 0.0002 | ± 0.0002 | ± 0.0002 | ± 0.0001 | |
| 27 | 0.7668 | 0.7438 | 0.7205 | 0.6928 | 0.6578 | 0.6121 | 0.5474 | 0.4409 | 0.6148 | 0.1983 | 37.079 |
| | ± 0.0001 | ± 0.0002 | ± 0.0002 | ± 0.0003 | ± 0.0003 | ± 0.0003 | ± 0.0004 | ± 0.0003 | ± 0.001 | ± 0.0001 | |